\documentclass[12pt]{article}
\newcommand{\Comment}[1]{{}}
\usepackage[textwidth = 450 pt, textheight = 630 pt]{geometry}
\usepackage[parfill]{parskip}
\usepackage{amssymb,euscript,amsmath,amsfonts}
\usepackage{tikz,tikz-cd,url}
\usetikzlibrary{decorations.pathreplacing,calligraphy}
\usepackage{xcolor}
\usepackage{graphicx}
\usepackage{color}
\usepackage{bbm}
\usepackage{tensor}
\usepackage{slashed}
\newcommand{\inv}[1]{\frac{1}{#1}}
\newcommand{\brac}[1]{ \left( #1 \right)}
\renewcommand{\cal}[1]{\mathcal{#1}}
\newcommand{\calz}{\cal{Z}}
\newcommand{\bcalz}{\bar{\cal{Z}}}

\newcommand{\bpsi}{\bar{\psi}}
\newcommand{\bchi}{\bar{\chi}}
\newcommand{\brho}{\bar{\rho}}
\newcommand{\bepsilon}{\bar{\epsilon}}
\newcommand{\hpsi}{\hat{\psi}}
\newcommand{\hbpsi}{\hat{\bpsi}}
\newcommand{\hchi}{\hat{\chi}}
\newcommand{\bbeta}{\Bar{\beta}}
\newcommand{\balpha}{\Bar{\alpha}}
\newcommand{\hrho}{\hat{\rho}}
\newcommand{\df}{\dot{f}}
\newcommand{\ddf}{\ddot{f}}
\newcommand{\hbchi}{\hat{\bchi}}
\newcommand{\hbrho}{\hat{\brho}}

\newcommand{\mn}{\mu\nu}

\newcommand{\bb}[1]{\mathbb{#1}}
\renewcommand{\frak}[1]{\mathfrak{#1}}
\newcommand{\bbm}[1]{\mathbbm{#1}}

\definecolor{MyDarkBlue}{rgb}{0.15,0.15,0.45}
\usepackage[linktocpage=true]{hyperref}
\hypersetup{
colorlinks=true,
citecolor=MyDarkBlue,
linkcolor=MyDarkBlue,
urlcolor=MyDarkBlue,
pdfauthor={Neil Lambert },
pdftitle={ },
pdfsubject={hep-th}
}

\usepackage{hypernat}
\usepackage{physics}
\usepackage{accents}
\usepackage{bm}
\usepackage[sorting=none]{biblatex}
\begin{document}

   \vspace{1.8truecm}

 \centerline{\Huge  {  Non-Relativistic Intersecting Branes, }} \vskip 12pt
 \centerline{\Huge  {Newton-Cartan Geometry   and AdS/CFT}}

\centerline{\LARGE \bf {\sc  }} \vspace{2truecm} \thispagestyle{empty} \centerline{
    {\large {{\sc Neil~Lambert}}}\footnote{E-mail address: \href{mailto:neil.lambert@kcl.ac.uk}{\tt neil.lambert@kcl.ac.uk} } and  {\large {{\sc Joseph~Smith}}}\footnote{E-mail address: \href{mailto:joseph.m.smith@kcl.ac.uk}{\tt joseph.m.smith@kcl.ac.uk} }
  }

\vspace{1cm}
\centerline{{\it Department of Mathematics}}
\centerline{{\it King's College London }} 
\centerline{{\it The Strand }} 
\centerline{{\it  WC2R 2LS, UK}} 

\vspace{1.0truecm}

\thispagestyle{empty}

\centerline{\sc Abstract}
\vspace{0.4truecm}
\begin{center}
\begin{minipage}[c]{360pt}{
    \noindent}

We discuss non-relativistic variants of four-dimensional ${\cal N=4}$ super-Yang-Mills theory  obtained from  generalised Newton-Cartan geometric limits of D3-branes in ten-dimensional spacetime. We argue that the natural interpretation of these limits is that they correspond to non-relativistic D1-branes or D3-branes intersecting the original D3-branes. The resulting gauge theories have dynamics that reduce to quantum mechanics on monopole moduli space or two-dimensional sigma-models on Hitchin moduli space respectively. We show that these theories possess interesting infinite-dimensional symmetries and we discuss the dual $AdS$ geometries.

\end{minipage}
\end{center}

\newpage
\tableofcontents

\section{Introduction}\label{sect: Intro}

There has  been a steady but growing interest in non-Lorentzian limits of String and M-theory (a selection of these paper is \cite{Gomis:2000bd,Gomis:2005pg,Bagchi:2009my,Andringa:2012uz,Harmark:2014mpa, Harmark:2017rpg,Roychowdhury:2022est,Bergshoeff:2022eog,Avila:2023aey,Bergshoeff:2023rkk}). Such limits consist of various generalizations of the classic non-relativistic limit of Einstein gravity which leads to so-called Newton-Cartan gravity associated to massive point particles. Since String and M-theory contain a variety of massive $p$-brane states one finds a corresponding variety of possible non-Lorentzian limits. As such these various limits are related to each other by a web of dualities that are inherited from the familiar dualities of String and M-theory \cite{Blair:2023noj}. To date these have mainly be applied to  supergravity theories, worldsheet string theories and Abelian $p$-brane actions.

In a recent paper \cite{Lambert:2024uue} we examined the  membrane-Newton-Cartan (MNC) limit of the M2-brane conformal field theory   and its associated $AdS_4\times S^7$ supergravity dual. The limit makes sense in the field theory leading to a novel non-Lorentzian field theory whose dynamics reduces to quantum mechanics on Hitchin moduli space. These theories have been constructed before \cite{Lambert:2018lgt,Lambert:2019nti} and shown to be maximally (or 3/4 maximally) supersymmetric. However one of the surprising features of the Lagrangian constructed from M2-branes  is that it admits an infinite dimensional spacetime symmetry group \cite{Lambert:2024uue}. That paper  also explored the gravitational dual, which is described by the MNC limit of eleven-dimensional supergravity constructed in \cite{Blair:2021waq} and was able to match the symmetries on both sides.  In this paper we wish to provide a similar analysis for the case of D3-branes, that is for  ${\cal N=4}$ super-Yang-Mills and its $AdS_5\times S^5$ dual (see also \cite{Fontanella:2024rvn} for another recent non-relativistic D3-brane  AdS/CFT construction).

To continue let us review the case of M-theory. Here there is a so-called membrane-Newton-Cartan  (MNC) limit where one re-scales time and and two space dimensions by a factor of $c$ and the remaining dimensions by $c^{-1/2}$  \cite{Blair:2021waq}.  We  can think of $c$, which is dimensionless, as controlling the speed of light. From the geometrical point of view this is encapsulated by a re-writing of the metric as
\begin{align}
     {\hat g}_{MNC}  = c^2\tau_{mn}dx^m\otimes dx^n + c^{-1} H_{mn}dx^m\otimes dx^n\ .
\end{align}
Here $m,n=0,1,2,...,10$ and $\tau_{mn}$ should   be thought of as a Lorentzian metric in three-dimensions whereas  $H_{mn}$ is a Riemannian metric in eight dimensions. However when viewed as eleven-dimensional tensors they are not individually invertable. Rather they represent a splitting of eleven-dimensional spacetime into a three-dimensional spacetime and eight dimensional transverse space. There is also a decomposition of the 3-form field in appropriate powers of $c$. 
For finite $c$  this is simply a coordinate transformation. However the point of this construction is that it is possible to take the limit $c\to\infty$ in such a way that one retains non-trivial dynamical equations.  This last condition determines the  curious power of $c^{-1/2}$ that is used to scale the remaining dimensions.

Reducing the above limit to type IIA String Theory one finds two possible limits, depending on whether the M-theory circle is taken to lie along the large directions, those contained in $\tau_{mn}$, or the small directions, those contained in $H_{mn}$ \cite{Blair:2021waq}.   Taking the M-theory circle along the large directions of $\tau_{mn}$ leads to the so-called String-Newton-Cartan (SNC) limit
\begin{align}
     {\hat g}_{SNC} &= c^2\tau_{\mu\nu}dx^\mu\otimes dx^\nu +  H_{\mu\nu} dx^\mu\otimes dx^\nu  \qquad e^{\hat \phi }= c e^\phi\ ,
\end{align}
where $\mu,\nu=0,1,2,...,9$ with $\tau_{\mu\nu}$  a two-dimensional Lorentzian metric and $H_{\mu\nu}$ an eight-dimensional Riemannian metric (note that we have redefined $c\to c^{2/3}$). The $c\to\infty$ limit is strongly coupled and hence this limit is somewhat formal: we are compactifying on a circle whose size is getting larger as $c\to\infty$. Nevertheless, when viewed within String Theory it exists as a limit of various configurations.

Alternatively, we could take the M-theory circle to lie in the small directions of $H_{mn}$ to find
\begin{align}
    {\hat g}_{D2NC} &= c^2\tau_{\mu\nu}dx^\mu\otimes dx^\nu + c^{-2} H_{\mu\nu} dx^\mu\otimes dx^\nu
\qquad e^{\hat \phi} = c^{-1} e^\phi\ ,
\end{align}
where $\tau_{\mu\nu}$ is three-dimensional and $H_{\mu\nu}$  seven-dimensional (and  we have redefined  $c\to c^{4/3}$). 
We refer to this as a D2-Newton-Cartan limit (D2NC).
Applying T-duality to this second case leads more generally to D$p$NC limits \cite{Blair:2023noj}:\footnote{In \cite{ Blair:2023noj} these were referred to as M$p$T limits.}
\begin{align} \label{eq: DpNC limit}
   {\hat g}_{DpNC} &= c^2\tau_{\mu\nu}dx^\mu\otimes dx^\nu +c^{-2} H_{\mu\nu} dx^\mu\otimes dx^\nu\qquad e^{\hat \phi} = c^{p-3}  e^\phi \ ,
\end{align}
where $\tau_{\mu\nu}$ is a $(p+1)$-dimensional Lorentzian metric and $H_{\mu\nu}$ a $(9-p)$-dimensional Euclidean metric.

A  feature of these constructions is that in order to cancel divergences as $c\to\infty$ one also needs a diverging $(p+1)$-form field. In particular, for the MNC limit discussed above one needs the 3-form field to have the form\footnote{Note that the divergent term here differs in sign from the discussion in \cite{Blair:2021waq}; the only significant effect of this is to flip the sign of the constraint, setting the self-dual sector of the totally transverse part of $F_4$ to zero.}
\begin{align}
    \hat{C}_3 = c^3 \tau^0 \wedge \tau^1 \wedge \tau^2 + C_3\ ,
\end{align}
with $C_3$ finite in the $c\to\infty$ limit. Using the map between the MNC and D2NC limits and T-dualising \cite{Blair:2023noj}, we find the D$p$NC limit requires the divergent structure 
\begin{align}\label{C-critical}
    \hat{C}_{p+1} = c^4 e^{-\varphi} \tau^0 \wedge ... \wedge \tau^p + C_{p+1} \ ,
\end{align}
in the RR $(p+1)$-form field, where we define $\varphi$ by $e^{\phi} = g_s e^{\varphi}$. A similar divergence in the Kalb-Ramond field is also required in the SNC limit \cite{Bergshoeff:2019pij}.

In the MNC solution of \cite{Lambert:2024uue} it was found that the divergent piece of the 3-form field is constant and therefore closed. We will see that this is also true for the limits of D3-branes that we shall consider in this paper, suggesting there is something deeper happening here. The presence of a constant form-field does nothing to the bulk supergravity equations of motion. Indeed, one might be tempted to simply  gauge it away. However, such gauge transformations are non-zero at infinity and thus act as asymptotic symmetries in the full String Theory. Furthermore there are $p$-brane states that are charged under these symmetries and therefore transform non-trivially under such a gauge transformation. In other words   a gauge transformation is only trivial if none of the objects present carry the associated charge. 

To see how these background fields arise physically, we can consider the D$p$-brane supergravity solution
\begin{subequations}
\begin{align}
    g &= H^{-1/2} \eta_{\mn} dx^{\mu}\otimes  dx^{\nu} + H^{1/2} \delta_{IJ} dX^I \otimes dX^J \ , \\
    C_{p+1} &=  H^{-1}   dt \wedge ... \wedge dx^p \ , \\
    e^{\phi} &= g_sH^{\frac{3-p}{4}} \ ,
\end{align}
\end{subequations}
where $H$ satisfies the equation
\begin{equation}
    \partial_I \partial_I H = 0 \ .
\end{equation}
Suppose we smear the brane over the transverse coordinates; then $H$ is constant, with this constant determining the asymptotic geometry of the solution\footnote{For this reason, we have neglected to include the usual subtracted constant term in the definition of $C_{p+1}$.}. In particular given the choice
\begin{equation}
    H = c^{-4} \ ,
\end{equation}
the solution becomes
\begin{subequations}
\begin{align}
    g &= c^2 \eta_{\mn} dx^{\mu}\otimes dx^{\nu} + c^{-2} \delta_{IJ} dX^I \otimes dX^J \ , \\
    C_{p+1} &= c^4 dt\wedge ... \wedge dx^p \ , \\
    e^{\phi} &= c^{p-3}g_s \ .
\end{align}
\end{subequations}
Taking $c\to \infty$ gives the D$p$NC limit of a flat background, complete with the correct coefficient of the divergent $(p+1)$-form field. As discussed in \cite{Blair:2023noj}, the bosonic part of the worldvolume theory on a stack of $N$ probe D$p$-branes aligned with the D$p$NC geometry in static gauge has the $c$-expansion
\begin{align} 
    S_p = - \frac{1}{2 g_{YM}^2} \tr \int d^{p+1} x \brac{ \inv{2} F_{\mn} F^{\mn} + D_{\mu} X^I D^{\mu} X^I - \inv{2} [X^I, X^J]^2 } + O\brac{c^{-4}} \ .
\end{align}
We have redefined the transverse coordinates by a factor of $2\pi \alpha'$ and defined the Yang-Mills coupling as
\begin{equation}
    g_{YM}^2 = \inv{\brac{2\pi\alpha'}^2 g_s T_p} \ .
\end{equation}
This is finite as we take $c\to \infty$, with the limit decoupling the higher-order terms in the worldvolume theory. The dynamics of the branes in the D$p$NC limit is governed by maximally supersymmetric $U(N)$ Yang-Mills in ($p+1$)-dimensions; the limit has made the low-energy approximation of the full DBI action exact.  It seems natural to expect that something similar happens in the MNC limit of M-Theory, with the dynamics of M2-branes aligned along the MNC limit reducing to that of the IR SCFT\footnote{In other words, the dynamics of two M2-branes in the MNC limit should be described by BLG, and the dynamics of a stack on an $\bb{Z}_k$ orbifold by ABJM.}. In particular we see that, when aligned,  the D$3$NC limit of the D3-brane CFT   and   MNC limit of the M2-brane CFT simply act as a symmetries.

The obvious next question is the fate of branes not aligned with the D$p$NC limit. Following the discussion above, we realise such limits using intersecting brane configurations where one of the branes implements the D$p$NC limit. Unlike in the aligned case, generically these set-ups will correspond to non-relativistic limits of the brane worldvolume theory. Engineering these limits using intersections of branes provides an easy way of seeing whether supersymmetry will be present in the non-relativistic field theory, which is hard to predict when directly working with the field theory.

In this way the set-up in   \cite{Lambert:2024uue} can be viewed as the configuration
\begin{align} 
\begin{array}{rrrrrr}
M2: & 0 & 1 &2&& \\
MNC:& 0 &  & &3&4 \ \ . \\
\end{array}
\end{align}
In particular, although the action is a non-relativistic three-dimensional gauge theory, the   dynamics restricts to quantum mechanics on Hitchin's moduli space with time being the only large dimension on the original M2-brane. This fits well with the interpretation of intersecting M2-branes as these are described  by a Hitchin system in the original worldvolume M2-brane CFT. 

In this paper we will explore such limits for D3 branes. The intersecting brane configurations we will consider are
\begin{align}
\begin{array}{rrrrrr}
D3: & 0 & 1 &2&3& \\
D1NC:& 0 &  & & &4 \ \ , \\
\end{array}
\end{align}
which implements the D1NC limit, and 
\begin{align}
\begin{array}{rrrrrrr}
D3: & 0 & 1 &2& 3 && \\
D3NC:& 0 & 1 & & &4&5 \ \ , \\
\end{array}
\end{align}
which implements the D3NC limit. The worldvolume theories in both cases correspond to different non-relativistic limits of $\cal{N}=4$ super Yang-Mills. The D1NC limit leads to quantum mechanics on monopole moduli space and the D3NC limit leads to a two-dimensional sigma-model on Hitchin moduli space. As in the M2 case, the dimensions of the sigma models comes from the number of large directions on the D3-brane and the dimensions of the soliton equations from the number of small directions. We will again find an infinite-dimensional extension of the spacetime symmetries.

The rest of this paper is organised as follows. In section \ref{sect: field theory} we will evaluate the D1NC and D3NC limits of four-dimensional $\cal N=4$ super-Yang-Mills and find the symmetries and associated conserved quantities. In section \ref{sect: gravity} we discuss how these field theories arise intersecting brane set-ups. This gives us non-relativistic brane solutions, which we can take the near-horizon limits of to find the corresponding limits of the dual $AdS_5\times S^5$ geometry. In section \ref{sect: relations} we discuss how the theories we obtain are related to each other and previously examined theories through string dualities. In section \ref{sect: conclusions} we give our conclusions. We also include an appendix discussing the null reduction of five-dimensional $\cal{N}=2$ super Yang-Mills, which gives the field theory that arises from the SNC limit of $\cal{N}=4$ that is S-dual to the D1NC limit. 

\section{Non-Relativistic Limits of \texorpdfstring{$\cal{N}=4$}{N=4} Super Yang-Mills} \label{sect: field theory}

\subsection{The D1NC Limit} \label{sect: D1NC}

\subsubsection{Action}

Let us start with the action for 4d $\cal{N}=4$ super-Yang-Mills in the form
\begin{align} \nonumber
   \hspace{-0.6cm} \hat{S} = \inv{2\hat{g}_{YM}^2} \tr \int d^4\hat{x} \bigg(& -\inv{2} \hat{F}_{\mn} \hat{F}^{\mn} - \hat{D}_{\mu} \hat{X} \hat{D}^{\mu} \hat{X}  - \hat{D}_{\mu} \hat{Y}^{M} \hat{D}^{\mu} \hat{Y}^M + \inv{2} [\hat{Y}^M, \hat{Y}^N]^2 \\ \label{eq: rel N=4 action 1}
    &+ [\hat{X}, \hat{Y}^M]^2 + i \hat{\bpsi} \Gamma^0 \Gamma^{\mu} D_{\mu} \hpsi + \hat{\bpsi} \Gamma^0 \Gamma_4 [\hat{X}, \hpsi] - \hat{\bpsi} \Gamma^0 \Gamma_5 \Gamma^M[ \hat{Y}^M, \hpsi]
    \bigg) \ .
\end{align}
Note that we've split one scalar field $\hat{X}$ off from the other five (indexed by $\{M,N,...\}$). The Fermion $\hat{\psi}$ is a real 32-component spinor satisfying the condition
\begin{equation}
    \Gamma_{012345} \hat{\psi} = - \hat{\psi} \ ,
\end{equation}
with $\{\Gamma^{\mu}, \Gamma^4, \Gamma^5 , \Gamma^M \}$ the gamma matrices for the real spinor representation of $SO(1,10)$. Throughout we will use a bar to denote conjugation of spinors, {\it i.e.} $\bpsi = \psi^{\dag}$.

We can then consider the coordinate scaling
\begin{subequations}
\begin{align}
    \hat{t} &= c t \ , \\
    \hat{x}^i &= c^{-1} x^i \ ,
\end{align}
\end{subequations}
which we note can be brought into the form of a more standard non-relativistic limit, {\it i.e.} one where only time is re-scaled, using the theory's conformal symmetry. In order to find a set of field scalings that lead to a non-trivial limit we must split $\hat{\psi}$ into chiral components with respect to $\Gamma_{05}$, {\it i.e.}
\begin{equation}
    \hat{\psi}_{\pm} = \inv{2} \brac{\bbm{1} \pm \Gamma_{05}} \hat{\psi} \ .
\end{equation}
We can then make the field redefinitions
\begin{subequations}
\begin{align}
    \hat{X}(\hat{t}, \hat{x}) &= c X(t,x) \ , \\
    \hat{Y}^M(\hat{t}, \hat{x}) &= c^{-1} Y^M(t,x) \ , \\
    \hat{A}_t(\hat{t}, \hat{x}) &= c^{-1} A_t(t,x) \ , \\
    \hat{A}_i(\hat{t}, \hat{x}) &= c A_i(t,x) \ , \\
    \hat{\psi}_+(\hat{t}, \hat{x}) &= c^{\frac{1}{2}} \psi_+(t,x) \ , \\
    \hat{\psi}_-(\hat{t}, \hat{x}) &= c^{- \frac{3}{2}} \psi_-(t,x) \ .
\end{align}
\end{subequations}
The powers of $c$ in the rescaling of the coordinates and scalar fields matches those of the D1NC limit of type IIB supergravity \eqref{eq: DpNC limit}. Following this, the action becomes
\begin{align} \nonumber
    \hat{S} = \inv{2c^2 \hat{g}_{YM}^2} \tr \int dt d^3x \bigg(& 
    -c^4 \brac{\tfrac12F_{ij} F_{ij} + D_i X D_i X} + F_{0i} F_{0i} + D_0 X D_0 X \\ \nonumber
    &- D_i Y^M D_i Y^M + [X,Y^M]^2 - i \bpsi_+ D_t \psi_+ - i \bpsi_+ \Gamma_{0i} D_i \psi_- \\
    &- i \bpsi_- \Gamma_{0i} D_i \psi_+ - 2 \bpsi_+ \Gamma_{04} [X, \psi_-] + \bpsi_+ \Gamma^M [Y^M, \psi_+] + O(c^{-4})
    \bigg) \ .
\end{align}
As we would like to keep the kinetic terms around, we should redefine the coupling as
\begin{equation}
    \hat{g}_{YM}^2 = \frac{g_{D1}^2}{c^2} \ ,
\end{equation}
with $g_{D1}$ finite. The D1NC limit is therefore a weakly-coupled limit of $\cal{N}=4$ SYM; again, this is consistent with the scaling of the dilaton in (\ref{eq: DpNC limit}). 

The divergent part of the action can be rewritten as
\begin{equation}
    \tr\brac{\tfrac12F_{ij} F_{ij} + D_i X D_i X} = \frac12\tr \brac{F_{ij} \mp \varepsilon_{ijk} D_k X}^2 \pm   \tr \brac{\varepsilon_{ijk} F_{ij} D_k X} \ ,
\end{equation}
with the Bianchi identity meaning the second term is a total derivative. We will see in section \ref{sect: D1NC brane setup} that this term is cancelled by the presence of a constant background 2-form field in the String Theory realisation of this limit. We can therefore introduce an antisymmetric Hubbard-Stratonovich auxiliary field $G_{ij}$ to rewrite it as
\begin{equation}
    S_+ = \inv{2g_{D1}^2} \tr \int dt d^3x \, \brac{ G_{ij} \brac{F_{ij} \mp \varepsilon_{ijk} D_k X} + \inv{4 c^4} G_{ij} G_{ij} } \ .
\end{equation}

As every term in the action is finite we can now take the $c\to \infty$ limit. Taking the upper sign in the divergent term, the action in the limit is
\begin{align} \label{eq: D1NCYM action} \nonumber
    S_{D1NC} = \inv{2g_{D1}^2} \tr \int dt d^3x \bigg(& F_{0i} F_{0i} + D_t X D_t X + G_{ij} \brac{F_{ij} - \varepsilon_{ijk} D_k X} \\ \nonumber
    &- D_i Y^M D_i Y^M + [X,Y^M]^2 - i \bpsi_+ D_t \psi_+  \\
    &- 2i \bpsi_- \Gamma_{0i} D_i \psi_+ - 2 \bpsi_- \Gamma_{04} [X, \psi_+] + \bpsi_+ \Gamma^M [Y^M, \psi_+] \bigg) \ .
\end{align}

This action was first constructed in \cite{Lambert:2018lgt} as the dimensional reduction of the five-dimensional theory that arises from M5-branes on a null circle, and can also be found by taking the non-relativistic limit of five-dimensional $\cal{N}=2$ super-Yang-Mills \cite{Lambert:2019nti}. Interpreting our coordinates and scalar fields as the spacetime coordinates of a D3-brane stack, we see from \eqref{eq: DpNC limit} that the limit corresponds to the D1NC limit of type IIB string theory.

\subsubsection{Bosonic Symmetries} \label{sect: D1NC symmetries}

Let us find the bosonic symmetries of the action \eqref{eq: D1NCYM action}. Starting with the spacetime symmetries, we find a preserved $\frak{so}(2,1)$ subalgebra of the relativistic conformal transformations that act on our coordinates as
\begin{subequations} \label{eq: D1NC time transformations}
\begin{align}
    \hat{t} &= t + f(t) \ , \\
    \hat{x}^i &= x^i \brac{1 + \df} \ , \\
    f(t) &= a + bt + ct^2 \ ,
\end{align}
\end{subequations}
where $a,b,c$ are constants, provided we take the fields to have the transformations
\begin{subequations}
\begin{align}
    \hat{X}(\hat{t},\hat{x}) &= \brac{1 - \df} X(t,x) \ , \\
    \hat{Y}^M(\hat{t}, \hat{x}) &= \brac{1 - \df} Y^M(t,x) \ , \\
    \hat{G}_{ij}(\hat{t}, \hat{x}) &= \brac{\brac{1 - 2 \df} G_{ij} - 2 \ddf F_{0[i} x_{j]} - \ddf \varepsilon_{ijk} x^k D_t X } (t,x)\ , \\
    \hat{A}_t(\hat{t}, \hat{x}) &= \brac{\brac{1 - \df} A_t - \ddf x^i A_i } (t,x) \ , \\
    \hat{A}_i(\hat{t}, \hat{x}) &= \brac{1 - \df} A_i(t,x) \ , \\
    \hpsi_+(\hat{t}, \hat{x}) &= \brac{1 - \frac{3}{2}\df} \psi_+(t,x) \ , \\
    \hpsi_-(\hat{t}, \hat{x}) &= \brac{\brac{1 - \frac{3}{2}\df}\psi_- + \frac{1}{2}\ddf x^i \Gamma_{0i} \psi_+} \ .
\end{align}
\end{subequations}
The conserved currents for these symmetries are\footnote{We will add improvement terms throughout to retain gauge-invariance.}
\begin{subequations}
\begin{align}
    0 &= \partial_0 j^{(a)}_0 + \partial_i j^{(a)}_i \ , \\ \nonumber
    j^{(a)}_0 &= \tr\Big(F_{0i} F_{0i} + D_tX D_t X - G_{ij} \brac{F_{ij} -\epsilon_{ijk} D_k X} \\ \nonumber
    & \hspace{1.15cm} + D_i Y^M D_i Y^M - [X,Y^M]^2 + 2i \bpsi_- \Gamma_{0i} D_i \psi_+ \\
    & \hspace{1.15cm}+ 2 \bpsi_- \Gamma_{04} [X,\psi_+] - \bpsi_+ \Gamma^M [Y^M, \psi_+] \Big) \ , \\
    j^{(a)}_i &= \tr\brac{2 G_{ij} F_{ij} + \epsilon_{ijk} G_{jk} D_t X - 2 D_t Y^M D_i Y^M - 2i \bpsi_- \Gamma_{0i} D_t \psi_+} \ ,
\end{align}
\end{subequations}
for $f= a$,
\begin{subequations}
\begin{align}
    0 &= \partial_0 j^{(b)}_0 + \partial_i j^{(b)}_i \ , \\ \nonumber
    j^{(b)}_0 &= \tr \bigg( F_{0i} \brac{t F_{0i} - x^j F_{ij}} + D_tX \brac{X + t D_t X + x^i D_i X} \\
    & \hspace{1.15cm} - \frac{i}{2} \bpsi_+ \brac{t D_t \psi_+ + x^i D_i \psi_+} - \inv{2} t \cal{L} \bigg) \ , \\ \nonumber
    j^{(b)}_i &= \tr \bigg( x^j F_{0i} F_{0j} + G_{ij} \brac{t F_{0j} + x^k F_{kj}} - \inv{2} \epsilon_{ijk} G_{jk} \brac{X + t D_t X + x^l D_l X} \\ \nonumber
    & \hspace{1.15cm} - D_i Y^M \brac{Y^M + t D_t Y^M + x^j D_j Y^M} - \inv{2} x^i \cal{L} \\
    & \hspace{1.15cm} - i \bpsi_- \Gamma_{0i} \brac{\frac{3}{2} \psi_+ + t D_t \psi_+ + x^j D_j \psi_+}  \bigg) \ ,
\end{align}
\end{subequations}
for $f= bt$, and
\begin{subequations}
\begin{align}
    0 &= \partial_0 j^{(c)}_0 + \partial_i j^{(c)}_i \ , \\ \nonumber
    j^{(c)}_0 &= \tr \bigg(
    t^2 F_{0i} F_{0i} - 2t x^j F_{0i} F_{ij} + 2t X D_t X + t^2 D_t X D_t X + 2t x^i D_t X D_i X \\
    & \hspace{1.15cm} - X^2 - \epsilon_{ijk} x^k X F_{ij} - \frac{i}{2} \bpsi_+ \brac{t^2 D_t \psi_+ + 2 t x^i D_i \psi_+} - \inv{2} t^2 \cal{L}
    \bigg) \ , \\ \nonumber
    j^{(c)}_i &= \tr \bigg(
    2t x^j F_{0i} F_{0j} + 2 \epsilon_{ijk} F_{0j} x^k + G_{ij} \brac{2t x^k F_{kj} + t^2 F_{0j}} - \epsilon_{ijk} G_{jk} \Big(tX \\ \nonumber
    & \hspace{1.15cm} + t x^j D_j X  + \inv{2} t^2 D_t X\Big) - D_i Y^M \Big(2t Y^M + 2t x^j D_j Y^M + t^2 D_t Y^M\Big)  \\
    & \hspace{1.15cm} - i \bpsi_- \Gamma_{0i} \brac{3t \psi_+ + 2t x^j D_j \psi_+ + t^2 D_t \psi_+} - t x^i \cal{L}
    \bigg) \ ,
\end{align}
\end{subequations}
for $f = c t^2$.

We can also consider time-dependent translations of the form
\begin{subequations} \label{eq: D1NC spatial translations}
\begin{align}
    \hat{t} &= t \ , \\
    \hat{x}^i &= x^i + \xi^i(t) \ ,
\end{align}
\end{subequations}
which are symmetries if the fields transform as
\begin{subequations}
\begin{align}
    \hat{X}(\hat{t},\hat{x}) &= X(t,x) \ , \\
    \hat{Y}^M(\hat{t}, \hat{x}) &= Y^M(t,x) \ , \\
    \hat{A}_t(\hat{t}, \hat{x}) &= \brac{A_t - \dot{\xi}^i A_i}(t,x) \ , \\
    \hat{A}_i(\hat{t}, \hat{x}) &= A_i(t,x) \ , \\
    \hat{G}_{ij}(\hat{t}, \hat{x}) &= \brac{G_{ij} - 2 F_{0[i} \dot{\xi}_{j]} - \epsilon_{ijk} \dot{\xi}_k D_t X - \epsilon_{ijk} \ddot{\xi}_k X  }(t,x) \ , \\
    \hpsi_+(\hat{t}, \hat{x}) &= \psi_+(t,x) \ , \\
    \hpsi_-(\hat{t}, \hat{x}) &= \brac{\psi_- + \inv{2} \dot{\xi}^i \Gamma_{0i} \psi_+}(t,x) \ .
\end{align}
\end{subequations}
The associated conserved current is
\begin{subequations}
\begin{align}
    0 &= \partial_i T_{ij} \ , \\ \nonumber
    T_{ij} &= \tr\bigg( F_{0i} F_{0j} + G_{ik} F_{jk} - D_i Y^M D_j Y^M - i \bpsi_- \Gamma_{0i} D_j \psi_+ \\ 
    &\hspace{1.26cm} - \inv{2}\epsilon_{ikl} G_{kl} D_j X 
    + \epsilon_{ijk} \partial_0\brac{X F_{0k} } + \delta_{ij} \partial_0^2\brac{X^2} - \inv{2} \delta_{ij} \cal{L} \bigg) \ ,
\end{align}
\end{subequations}
which has no timelike component.

In contrast, the $\frak{so}(3)$ algebra of spatial rotations
\begin{subequations} \label{eq: D1NC rotations}
\begin{align}
    \hat{t} &= t \ , \\
    \hat{x}^i &= x^i + \omega_{ij} x^j \ , \\
    \omega_{ij} &= - \omega_{ji} \ ,
\end{align}
\end{subequations}
cannot be made time-dependent, unlike in the M2 limit previously considered in \cite{Lambert:2024uue}. This is a symmetry with the field transformations
\begin{subequations} \label{eq: rotations field transformations}
\begin{align}
    \hat{A}_t(\hat{t},\hat{x}) &= A_t(t,x) \ , \\
    \hat{A}_i(\hat{t}, \hat{x}) &= \brac{A_i + \omega_{ij} A_j}(t,x) \ , \\
    \hat{X}(\hat{t},\hat{x}) &= X(t,x) \ , \\
    \hat{Y}^M(\hat{t}, \hat{x}) &= Y^M(t,x) \ , \\
    \hat{G}_{ij}(\hat{t}, \hat{x}) &= \brac{G_{ij} + \omega_{ik} G_{kj} + \omega_{jk} G_{ik}}(t,x) \ , \\
    \hpsi_{\pm}(\hat{t}, \hat{x}) &= \brac{1 + \inv{4} \omega_{ij} \Gamma_{ij}} \psi_{\pm}(t,x) \ , 
\end{align}
\end{subequations}
with conserved current
\begin{subequations}
\begin{align}
    0 &= \partial_0 M_{0ij} + \partial_k M_{kij} \ , \\
    M_{0ij} &= \tr \brac{
    F_{0k} F_{k[i} x_{j]} - D_t X D_{[i} X x_{j]} - \frac{i}{8} \bpsi_+ \Gamma_{ij} \psi_+ + \frac{i}{2} \bpsi_+ D_{[i} \psi_+ x_{j]} } \ , \\ \nonumber
    M_{kij} &= \tr \bigg(
     - F_{0k} F_{0[i} x_{j]} +  G_{kl} F_{l[i} x_{j]} + \inv{2} \epsilon_{klm} G_{lm} D_{[i} X x_{j]} + D_k Y^M D_{[i} Y^M x_{j]} \\
     & \hspace{1.3cm} - \frac{i}{4} \bpsi_- \Gamma_{0k} \Gamma_{ij} \psi_+ + i \bpsi_- \Gamma_{0k} D_{[i} \psi_+ x_{j]} + \inv{2} \delta_{k[i} x_{k]} \cal{L} \bigg) \ .
\end{align}
\end{subequations}

Next, we turn to the R-symmetry. Taking the non-relativistic limit of the scalar fields breaks the original $\frak{so}(6)_R$ transformations to an $\frak{so}(5)_R$ that rotate the $Y^M$ fields into themselves, and the non-relativistic avatar of the transformations that mix the two types of scalars. Let us deal with the $\frak{so}(5)_R$ transformations first. The action is invariant under the transformation\footnote{From here onwards we omit any fields that transform trivially.}
\begin{subequations} \label{eq: SO(5) R symmetry}
\begin{align}
    \hat{Y}^M(t,x) &= \brac{Y^M + r^{MN} Y^N}(t,x) \ , \\
    \hpsi_{\pm}(t,x) &= \brac{1 + \inv{4} \Gamma^{MN} r^{MN}}\psi_{\pm}(t,x) \ , 
\end{align}
\end{subequations}
with $r^{MN} = - r^{NM}$, with the corresponding conserved current
\begin{subequations}
\begin{align}
    0 &= \partial_0 J^{MN}_0 + \partial_i J^{MN}_i \ , \\
    J^{MN}_0 &= - \frac{i}{4} \tr \brac{\bpsi_+ \Gamma^{MN}\psi_+} \ , \\
    J^{MN}_i &= \tr\brac{Y^M D_i Y^N - Y^N D_i Y^M - \frac{i}{2} \bpsi_- \Gamma_{0i} \Gamma^{MN} \psi_+} \ .
\end{align}
\end{subequations}
The non-relativistic limit of the R-symmetry transformations that mix $X$ and $Y^M$ become a field-space analogue of Galilean boosts, with the transformations
\begin{subequations}
\begin{align}
    \hat{Y}^M(t,x) &= \brac{Y^M + v^M X}(t,x) \ , \\
    \hat{G}_{ij}(t,x) &= \brac{G_{ij} - \epsilon_{ijk} v^M D_k Y^M}(t,x) \ , \\
    \hpsi_-(t,x) &= \brac{\psi_- + \inv{2} \Gamma_{04} \Gamma^M v^M \psi_+}(t,x) \ , 
\end{align}
\end{subequations}
leaving the action invariant for any time-dependent $v^M$. The spatial conserved current is
\begin{subequations}
\begin{align}
    0 &= \partial_i j_i^M \ , \\
    j_i^M &= \tr \brac{2 X D_i Y^M - \epsilon_{ijk} Y^M F_{jk} + \inv{2} \bpsi_+ \Gamma_{i4} \Gamma^M \psi_+} \ .
\end{align}
\end{subequations}

Finally, we note that the transformation
\begin{subequations} \label{eq: strange transformation}
\begin{align}
    \hat{A}_t(t,x) &= \brac{A_t + \chi X}(t,x) \ , \\
    \hat{G}_{ij}(t,x) &= \brac{G_{ij} - \chi \epsilon_{ijk} F_{0k}}(t,x) \ , \\
    \hpsi_-(t,x) &= \brac{\psi_- - \inv{2} \chi \Gamma_{04} \psi_+}(t,x) \ ,
\end{align}
\end{subequations}
leaves the action invariant up to a total derivative, with conserved current
\begin{subequations}
\begin{align}
    0 &= \partial_0 \cal{J}_0 + \partial_i \cal{J}_i \ , \\
    \cal{J}_0 &= \epsilon_{ijk} \tr \brac{A_i \partial_j A_k - \frac{2i}{3} A_i A_j A_k} \ , \\
    \cal{J}_i &= - \tr\brac{
    \epsilon_{ijk} \brac{A_k \partial_k A_0 - A_j \partial_0 A_k +  A_0 \partial_i A_j - 2i A_0 A_i A_j} + 2 X F_{0i} - \frac{i}{2} \bpsi_+ \Gamma_{i4} \psi_+ } \ .
\end{align}
\end{subequations}
While the conservation equation is gauge-invariant, the components of the current are not. However, the conserved charge
\begin{equation}
    Q = \epsilon_{ijk} \tr \int_{\bb{R}^3} d^3x \brac{A_i \partial_j A_k - \frac{2i}{3} A_i A_j A_k} 
\end{equation}
\textit{is} gauge-invariant under transformations that are trivial at spatial infinity.

\subsubsection{Supersymmetry}

In \cite{Lambert:2018lgt} it was shown that the action \eqref{eq: D1NCYM action} is supersymmetric. However, the nature of the symmetries ({\it i.e.} whether they are physical or not) was not ascertained.  As spatial translations can be given arbitrary time-dependence while time translations remain physical, it is natural to predict that half our relativistic supersymmetry becomes time-dependent (and are therefore a redundancy of the description) and only half remains physical; we will show that this is indeed the case. While the extension of this analysis to the full set of relativistic superconformal transformations is clearly of interest, we will not pursue this here.

The spinor parameter in the relativistic case is a real 32-component spinor satisfying the condition
\begin{equation}
    \Gamma_{012345} \epsilon =  \epsilon \ .
\end{equation}
As with the fermions, after taking the non-relativistic limit it will be most straightforward to split this into its chiral components with respect to $\Gamma_{05}$. This leaves us with two spinor parameters $\epsilon_{\pm}$ for which
\begin{subequations}
\begin{align}
     \Gamma_{012345} \epsilon_{\pm} = - \epsilon_{\pm} \ , \\
     \Gamma_{05} \epsilon_{\pm} = \pm \epsilon_{\pm} \ .
\end{align}
\end{subequations}

Let us deal with $\epsilon_-$ first. Then, the dimensional reduction of the transformations in \cite{Lambert:2018lgt}\footnote{Note that we use slightly different spinor normalisations than those appearing in that work.} give the supersymmetry transformations
\begin{subequations}
\begin{align}
    \delta X &= 0 \ , \\
    \delta Y^M &= -i \bepsilon_- \Gamma_{0M} \psi_+ \ , \\
    \delta A_0 &= - i \bepsilon_- \Gamma_0 \psi_+ \ , \\
    \delta A_i &= 0 \ , \\
    \delta G_{ij} &= i \bepsilon_- \brac{\Gamma_k \Gamma_{ij} D_k \psi_- -i \Gamma_4 \Gamma_{ij} [X,\psi_-]} - i \partial_0 \bepsilon_- \Gamma_0 \Gamma_{ij} \psi_+  \ , \\
    \delta \psi_+ &=  \frac{1}{2} \Gamma_0 \Gamma_{ij} \brac{F_{ij} + \epsilon_{ijk} D_k X} \epsilon_- \ , \\
    \delta \psi_- &= - \brac{ F_{0i} \Gamma_i + D_0 X \Gamma_4 +i [X,Y^M] \Gamma_4 \Gamma^M - D_i Y^M \Gamma_i \Gamma^M} \epsilon_-  - 2 \Gamma_4 X \partial_0 \epsilon_- \ ,
\end{align}
\end{subequations}
where we have allowed our spinor parameter to have arbitrary time-dependence, $\epsilon_- = \epsilon_-(t)$, and added a term proportional to its derivative in the transformations of $G_{ij}$ and $\psi_-$ to compensate. It is straightforward to see that the one-fermion terms in the variation cancel. Cancellation of the three-fermion terms requires the use of Fierz identities; however, as noted in \cite{Lambert:2018lgt}, a quick way to see that they must cancel is to note that they are contained within the three-fermion terms of the original relativistic theory, which we know is supersymmetric. 

The supercurrent associated with the transformation is
\begin{subequations}
\begin{align}
    0 &= \partial_i S^i \ , \\ \nonumber
    S^i &= \tr\Big(\brac{i F_{0j} \Gamma_j + i \Gamma_4 D_0 X - i \Gamma_{jM} D_j Y^M + \Gamma_{4M} [X,Y^M] }\Gamma_{0i} \psi_+ \\
    &\hspace{0.7cm} + i \Gamma_{ijk} F_{jk} \psi_- - 2 i \epsilon_{ijk} \Gamma_j \psi_- D_k X - 2 i \Gamma_{0i4} D_0 \brac{X \psi_+} \Big)  \ .
\end{align} 
\end{subequations}
The time-dependence of $\epsilon_-$ means this has no timelike component and there is no codimension-one conserved supercharge. 

We now consider $\epsilon_+$. The transformations which leave the action invariant are
\begin{subequations}
\begin{align}
    \delta X &= i \bepsilon_+ \Gamma_4 \psi_+ \ , \\
    \delta Y^M &= - i \bepsilon_+ \Gamma_{0M} \psi_+ \ , \\
    \delta A_0 &= i \bepsilon_+ \Gamma_0 \psi_- \ , \\
    \delta A_i &= i \bepsilon_+ \Gamma_i \psi_+ \ , \\
    \delta G_{ij} &= i \bepsilon_+ \brac{\Gamma_0 \Gamma_{ij} D_0 \psi_- + \Gamma_{ij} \Gamma_{0M} [Y^M, \psi_-] } \ , \\
    \delta \psi_+ &= F_{0i} \Gamma_i \epsilon_+ + D_0 X \Gamma_4 \epsilon_+ + D_i Y^M \Gamma_{iM} \epsilon_+ - i [X,Y^M] \Gamma_{4M} \epsilon_+ \ , \\
    \delta \psi_- &= - D_0 Y^M \Gamma_{0M} \epsilon_+ + \frac{i}{2} [Y^M, Y^N] \Gamma_0 \Gamma_{MN} \epsilon_+ + \frac{1}{2} G_{ij} \Gamma_0 \Gamma_{ij} \epsilon_+ \ ,
\end{align}
\end{subequations}
with corresponding conserved supercurrent
\begin{subequations}
\begin{align}
    0 &= \partial_0 \cal{S}^0 + \partial_i \cal{S}^i \ , \\
    \cal{S}^0 &= \tr\Big(\brac{F_{0i} \Gamma_i + D_0 X \Gamma_4 - D_i Y^M \Gamma_{iM} + i [X,Y^M] \Gamma_{4M} } \psi_+ \Big) \ , \\ \nonumber
    \cal{S}^i &=  \tr\bigg(\brac{D_0 Y^M \Gamma_{iM} - \frac{i}{2}[Y^M, Y^N] \Gamma_i \Gamma_{MN}  - \inv{2} \Gamma_{jk} \Gamma_i G_{jk} } \psi_+  + \Big( D_0 X \Gamma_{0i4} \\
    & \hspace{0.7cm} - F_{0j} \Gamma_0 \Gamma_j \Gamma_i +  + D_j Y^M \Gamma_{0M} \Gamma_j \Gamma_i  + i [X, Y^M] \Gamma_{0M} \Gamma_{i4}  \Big)  \bigg) \psi_-  \ ,
\end{align}
\end{subequations}
after using an improvement term to subtract a contribution proportional to the equation of motion of $G_{ij}$. As the timelike component is now non-trivial, we see that this is a physical symmetry and cannot be made time-dependent, as expected.

\subsection{The D3NC Limit} \label{eq: D3NC Limit}

\subsubsection{Action} \label{sect: D3NC action}

We can find another limit by starting with the $\cal{N}=4$ action in the form
\begin{align} \nonumber
    \hat{S} = \inv{2\hat{g}_{YM}^2} \tr \int d^4\hat{x} \bigg(& -\inv{2} \hat{F}_{\mn} \hat{F}^{\mn} - \hat{F}_{\mu i} \hat{F}^{\mu i} - \inv{2} \hat{F}_{ij} \hat{F}_{ij} - \hat{D}_{\mu} \hat{X}^a \hat{D}^{\mu} \hat{X}^a \\ \nonumber
    &- \hat{D}_i \hat{X}^a \hat{D}_i \hat{X}^a - \hat{D}_{\mu} \hat{Y}^{A} \hat{D}^{\mu} \hat{Y}^A - \hat{D}_i \hat{Y}^A \hat{D}_i \hat{Y}^A \\ \nonumber 
    &+ \inv{2} [\hat{Y}^A, \hat{Y}^B]^2  + [\hat{X}^a, \hat{Y}^A]^2
    + \inv{2}[\hat{X}^a, \hat{X}^b]^2 
    + i \hbpsi \Gamma^0 \Gamma^{\mu} \hat{D}_{\mu} \hpsi \\ \label{eq: rel N=4 action 2}
    &+ i \hbpsi \Gamma^0 \Gamma^i \hat{D}_i \hpsi + \hbpsi \Gamma^0 \Gamma_a [\hat{X}^a, \hpsi] + \hbpsi \Gamma^0 \Gamma^A [\hat{Y}^A,\hpsi] \bigg) \ ,
\end{align}
where we have split our indices into the groupings $\mu = 0,1$, $i=2,3$, $a=4,5$, $A=6,7,8,9$. Here $\{\Gamma^{\mu}, \Gamma^i, \Gamma^a, \Gamma^A\}$ is a real representation of $SO(1,9)$, with $\hat{\psi}$ a real 32-component spinor satisfying
\begin{equation}
    \Gamma_{01...9} \hat{\psi} = \hat{\psi} \ .
\end{equation}

Let's deal with the bosonic part of the action first. Taking the 
coordinates to have the scaling
\begin{subequations}
\begin{align}
    \hat{\sigma}^{\mu} &= c \sigma^{\mu} \ , \\
    \hat{x}^i &= c^{-1} x^i \ ,
\end{align}
\end{subequations}
we can propose the field redefinitions
\begin{subequations}
\begin{align}
    \hat{X}^a(\hat{\sigma},\hat{x}) &= c X^a(\sigma, x) \ , \\
    \hat{Y}^A(\hat{\sigma},\hat{x}) &= c^{-1} Y^A(\sigma, x) \ , \\
    \hat{A}_{\mu}(\hat{\sigma},\hat{x}) &= c^{-1} A_{\mu}(\sigma, x) \ , \\
    \hat{A}_i(\hat{\sigma},\hat{x}) &= c A_i(\sigma, x) \ .
\end{align}
\end{subequations}
Similarly to the limit considered in the previous section, the powers of $c$ here match the D3NC limit of type IIB supergravity \eqref{eq: DpNC limit}. This leaves us with the bosonic action
\begin{align} \nonumber
    \hat{S}_B = -\inv{2\hat{g}_{YM}^2} \tr \int d^2\sigma d^2x \Bigg[& 
     c^4 \brac{F_{23}^2 + D_i X^a D_i X^a - \inv{2} [X^a, X^b]^2} + F_{\mu i} F^{\mu i} \\ \nonumber
    &+ D_{\mu} X^a D^{\mu} X^a + D_i Y^A D_i Y^A - [X^a, Y^A]^2 \\
    &- c^{-4} \brac{ F_{01}^2 - D_{\mu} Y^A D^{\mu} Y^A + \inv{2} [Y^A, Y^B]^2 }
    \Bigg] \ . \label{eq: D3NC partial action}
\end{align}
To simplify notation we define $F \equiv F_{23}$. As we have no overall powers of $c$ in the action, we take the coupling to have no scaling,
\begin{equation}
    \hat{g}_{YM} = g_{D3} \ ,
\end{equation}
in agreement with (\ref{eq: DpNC limit}).

It will be convenient to combine the two 'large' scalars into the complex field
\begin{equation}
    \calz = X^4 + i X^5 \ .
\end{equation}
The divergent part of the action can then be written as
\begin{align} \nonumber
    S_+ &= - \frac{c^4}{2g_{D3}^2} \tr \int d^2 \sigma d^2x \brac{F^2 + \inv{4}[\calz, \bcalz]^2 + D_i \calz D_i \bcalz} \\
    &= - \frac{c^4}{2g_{D3}^2} \tr \int d^2\sigma d^2x \brac{
    \brac{F \pm \inv{2} [\calz, \bcalz]}^2 \mp F[\calz, \bcalz] + D_i \calz D_i \bcalz } \ .
\end{align}
Using the identity
\begin{equation}
    \tr\brac{F [\calz, \bcalz]} = i \tr \brac{\bcalz \brac{D_2 D_3 - D_3 D_2} \calz} \ ,
\end{equation}
we see that we can rewrite the final two terms as
\begin{align}
    \tr \brac{ D_i \calz D_i \bcalz \mp F[\calz, \bcalz] } =&\tr \brac{\brac{D_2 \pm i D_3} \calz \brac{D_2 \mp i D_3} \bcalz}\nonumber\\
    &\pm i \left(\partial_2\tr (\calz D_3 \bcalz  ) -\partial_3\tr (\calz   D_2 \bcalz  )  \right). \label{eq: D3NC total derivative bit}
\end{align}
The second line is a total derivative that, as for the D1NC limit, will be cancelled when we embed the limit in String Theory, as we shall discuss in section \ref{sect: D3NC brane setup}.

Introducing complex coordinates $z = x^2 + i x^3$, we can take the upper sign to get
\begin{equation}
    S_+ = - \frac{c^4}{2g_{D3}^2} \tr \int d^2 \sigma d^2 x \brac{ \brac{F + \inv{2} [\calz, \bcalz]}^2 + \Bar{D} \calz D \bcalz} \ .
\end{equation}
This is the sum of two squared quantities, so introducing two Hubbard-Stratonovich fields $B$ and $H$ allows the action to be rewritten in the form
\begin{align}
    S_+ = - \inv{2g_{D3}^2} \tr \int d^2 \sigma d^2x \brac{
    B \brac{F + \inv{2}[\calz, \bcalz] } + \bar{H} \bar{D} \calz + H D \bcalz - \inv{c^4} \brac{\inv{2} B^2 + H \bar{H}}
    } \ .
\end{align}
We can now take the $c\to\infty$ limit to get
\begin{align} \nonumber
    S_{D3NC,B} = \inv{2g_{D3}^2} \tr \int d^2 \sigma d^2 x \Bigg( & 8 \brac{F_{-z} F_{+\bar{z}} + F_{+z} F_{- \bar{z}}} - B \brac{F + \inv{2} [\calz, \bcalz]} \\ \nonumber
    &+ 2  \big( D_+ \calz D_- \bcalz + D_- \calz D_+ \bcalz \big) - H \Bar{D} \calz \\ \label{eq: D3NCYM b action}
    &- \bar{H} D \bcalz - 4 D Y^A \bar{D} Y^A + [\calz,Y^A] [\bcalz, Y^A] \Bigg)  \ ,
\end{align}
where we've written the action in terms of lightcone coordinates
\begin{equation}
    \sigma^{\pm} = \sigma^0 \pm \sigma^1\ ,
\end{equation}
and complex coordinates in the two planes. This theory was first constructed in \cite{Bershadsky:1995vm} after performing a partial topological twist of the relativistic theory: from our perspective, the topological twist is naturally implemented by taking the non-relativistic limit.

We now turn our attention to the fermions. In order to take the non-relativistic limit we must first identify a sensible way to split our spinors, with the two component scaling differently with $c$. We do this by determining the matrix $\gamma$ such that the BPS equations we derive upon taking the supersymmetry parameter $\epsilon$ to be an eigenvector of $\gamma$ with positive eigenvalue are the constraint equations we land on when taking the non-relativistic limit of the bosonic sector. In our case, this fixes $\gamma = \Gamma_{2345}$. Let us therefore define the fermion components
\begin{subequations}
\begin{align}
    \hat{\rho} &=  \inv{2} \brac{\bbm{1} + \Gamma_{2345}}\hat{\psi} \ , \\
    \hat{\chi} &= \inv{2} \brac{\bbm{1} - \Gamma_{2345}} \hat{\psi} \ .
\end{align}
\end{subequations}
It will be convenient to further split these into chiral components with respect to $\Gamma_{01}$,
\begin{subequations}
\begin{align}
    \hat{\chi}_{\pm} &= \inv{2} \brac{\bbm{1} \pm \Gamma_{01}} \hat{\chi} \ , \\
    \hat{\rho}_{\pm} &= \inv{2} \brac{\bbm{1} \pm \Gamma_{01}} \hat{\rho} \ .
\end{align}
\end{subequations}
In these variables the fermion action is
\begin{align} \nonumber
    \hat{S}_F = -\frac{1}{\hat{g}_{YM}^2} \tr \int d^4\hat{x} \bigg[& i \hbchi_+ \hat{D}_+ \hchi_+ + i\hbchi_- \hat{D}_- \hchi_- + i \hbrho_+ \hat{D}_+ \hrho_+ + i\hbrho_- \hat{D}_- \hrho_- \\ \nonumber
    &+ i \hbchi_- \Gamma_{0i} \hat{D}_i \hrho_+ + i \hbchi_+ \Gamma_{0i} \hat{D}_i \hrho_- + \hbchi_+ \Gamma_{0a} [\hat{X}^a, \hrho_-] \\
    &+ \hbchi_- \Gamma_{0a} [\hat{X}^a, \hrho_+] + \hbchi_+ \Gamma_{0A} [\hat{Y}^A, \hchi_-] + \hbrho_+ \Gamma_{0A} [\hat{Y}^A, \hrho_-] \bigg] \ .
\end{align}
We can then take the $c\to\infty$ limit with the scaling
\begin{subequations}
\begin{align}
    \hat{\rho}_{\pm}(\hat{\sigma}, \hat{x}) &= c^{\frac{1}{2}} \rho_{\pm}(\sigma,x) \ , \\
    \hat{\chi}_{\pm}(\hat{\sigma}, \hat{x}) &= c^{-\frac{3}{2}} \chi_{\pm}(\sigma,x) \ ,
\end{align}
\end{subequations}
to get
\begin{align} \nonumber
    S_{D3NC,F} = - \frac{1}{g_{D3}^2} \tr \int d^2\sigma d^2x \bigg[ & i \brho_+ D_+ \rho_+ + i \brho_- D_- \rho_- + 2i \bchi_- \brac{\Gamma_{0\bar{z}} D + \Gamma_{0 z} \bar{D} } \rho_+ \\ \nonumber
    &+ 2 i \bchi_+ \brac{\Gamma_{0\bar{z}} D + \Gamma_{0 z} \bar{D} } \rho_- + \bchi_+ \Gamma_{0\calz} [\calz, \rho_-] + \bchi_+ \Gamma_{0\bcalz} [\bcalz, \rho_-] \\ \label{eq: D3NCYM f action}
    &+ \bchi_- \Gamma_{0\calz} [\calz, \rho_+] + \bchi_- \Gamma_{0\bcalz} [\bcalz, \rho_+] + \brho_+ \Gamma_{0A} [Y^A, \rho_-] \bigg] \ ,
\end{align}
where we've defined
\begin{subequations}
\begin{align}
    \Gamma_z &= \inv{2} \brac{\Gamma_2 - i \Gamma_3} \ , \\
    \Gamma_{\calz} &= \inv{2} \brac{\Gamma_4 - i \Gamma_5} \ .
\end{align}
\end{subequations}

\subsubsection{Bosonic Symmetries} \label{sect: D3NC bosonic symmetries}

Let us find the bosonic symmetries of the actions \eqref{eq: D3NCYM b action} and \eqref{eq: D3NCYM f action}. Starting with the spacetime symmetries of the $\sigma$-directions, we find an enhancement of the expected $\frak{so}(2,2)$ symmetry algebra to two copies of the Virasoro algebra,
\begin{subequations}
\begin{align}
    \hat{\sigma}^{\pm} &= \sigma^{\pm} + f^{\pm}(\sigma^{\pm}) \ , \\
    \hat{z} &= z \ ,
\end{align}
\end{subequations}
provided we take the fields to have the transformation rules
\begin{subequations}
\begin{align}
    \hat{\calz}(\hat{\sigma}, \hat{x}) &= \calz(\sigma,x) \ , \\
    \hat{Y}^A(\hat{\sigma}, \hat{x}) &= \brac{1 - \inv{2}\partial_{+}f^+ - \inv{2} \partial_- f^- } Y^A(\sigma, x) \ , \\
    \hat{B}(\hat{\sigma}, \hat{x}) &= \brac{1 - \partial_+ f^+ - \partial_- f^-} B(\sigma, x) \ , \\
    \hat{H}(\hat{\sigma}, \hat{x}) &= \brac{1 - \partial_+ f^+ - \partial_- f^-} H(\sigma, x) \ , \\
    \hat{A}_{\pm}(\hat{\sigma}, \hat{x}) &= \brac{1 - \partial_{\pm} f^{\pm}} A_{\pm}(\sigma, x) \ , \\
    \hat{A}_z(\hat{\sigma}, \hat{x}) &= A_z(\sigma, x) \ , \\
    \hrho_{\pm}(\hat{\sigma}, \hat{x}) &= \brac{1 - \inv{2} \partial_{\mp} f^{\mp}} \rho_{\pm}(\sigma, x) \ , \\
    \hchi_{\pm}(\hat{\sigma}, \hat{x}) &= \brac{1 - \inv{2} \partial_{\pm} f^{\pm} - \partial_{\mp} f^{\mp}} \chi_{\pm}(\sigma, x) \ .
\end{align}
\end{subequations}
In particular, note that the $Y^A$ fields are no longer scalars under this transformation. The associated conserved currents take the form
\begin{subequations}
\begin{align}
    0 &= \partial_{\mp} \cal{T}^{\pm} + \partial \cal{T}^{\pm, z} + \bar{\partial} \cal{T}^{\pm, \bar{z}} \ , \\
    \cal{T}^{\pm} &= \tr \brac{16 F_{\pm z} F_{\pm \bar{z}} + 4 D_{\pm} \calz D_{\pm} \bcalz - 2i \brho_{\mp} D_{\pm} \rho_{\mp} } \ , \\ \nonumber
    \cal{T}^{\pm, z} &= \tr \Big(8 F_{\pm \bar{z}} F_{\mp \pm} + 2i B F_{\pm \bar{z}} - \bar{H} D_{\pm} \bcalz + 2 Y^A D_{\pm} \bar{D} Y^A \\
    & \hspace{1.15cm }- 2 D_{\pm} Y^A \bar{D} Y^A - 4i \bchi_{-} \Gamma_{0\bar{z}} D_{\pm} \rho_+ - 4 i \bchi_{+} \Gamma_{0\bar{z}} D_{\pm} \rho_-  \Big) \ .
\end{align}
\end{subequations}

As with the M2-brane  limit discussed in \cite{Lambert:2024uue}, in this theory the spatial symmetries are  enhanced: we find that transformations of the form
\begin{subequations}
\begin{align}
    \hat{z} &= z + f(z,\sigma) \ , \\
    \hat{\sigma}^{\pm} &= \sigma^{\pm} \ ,
\end{align}
\end{subequations}
where $f$ is a holomorphic function of $z$ with arbitrary dependence on the $\sigma^{\pm}$ coordinates, are symmetries of the theory. The fields have the transformations
\begin{subequations}
\begin{align}
    \hat{\calz}(\hat{\sigma}, \hat{x}) &= \brac{1 - \partial f} \calz(\sigma, x) \ , \\
    \hat{Y}^A(\hat{\sigma}, \hat{x}) &= Y^A(\sigma, x) \ , \\
    \hat{B}(\hat{\sigma}, \hat{x}) &= \brac{B + 2i \eta^{\alpha \beta}  \brac{\partial_{\alpha}f F_{\beta z} - \partial_{\alpha} \bar{f} F_{\beta \bar{z}} }}(\sigma, x) \ , \\
    \hat{H}(\hat{\sigma}, \hat{x}) &= \brac{H - 4 \partial_{+} \bar{f} D_- \bcalz - 4 \partial_- \bar{f} D_+ \bcalz - 4 \partial_+ \partial_- \bar{f} \bcalz}(\sigma,z) \ , \\
    \hat{A}_{\pm}(\hat{\sigma}, \hat{x}) &= \brac{A_{\pm} - \partial_{\pm} f A_z - \partial_{\pm} \bar{f} A_{\bar{z}}}(\sigma, x) \ , \\
    \hat{A}_z(\hat{\sigma}, \hat{x}) &= \brac{1- \partial f} A_z(\sigma, x) \ , \\
    \hrho_{\pm}(\hat{\sigma}, \hat{x}) &= \brac{1 - \inv{2}\brac{\partial f + \bar{\partial} \bar{f}} + \frac{i}{2}\Gamma_{23} \brac{\partial f - \bar{\partial} \bar{f}}} \rho_{\pm}(\sigma, x)  \ , \\
    \hchi_{\pm}(\hat{\sigma}, \hat{x}) &= \Big(\chi_{\pm} +\brac{\partial_{\mp} f \Gamma_{0z} + \partial_{\mp} \bar{f} \Gamma_{0\bar{z}} }\rho_{\mp} \Big) (\sigma, x) \ ,
\end{align}
\end{subequations}
under the symmetry, and the corresponding conserved current is
\begin{subequations}
\begin{align}
    0 &= \bar{\partial} T \ , \\ \nonumber
    T &= \tr \Big(16 F_{+z} F_{-z} + \calz DH - 4 D Y^A D Y^A \\
    & \hspace{1.15cm}+ 4i D \bchi_- \Gamma_{0\bar{z}} \rho_+ + 4i D \bchi_+ \Gamma_{0\bar{z}} \rho_-  \Big) \ .
\end{align}
\end{subequations}

We now move on to the internal symmetries. The split in the scalar fields when performing the $c\to \infty$ limit means we expect the original $\frak{so}(6)_R$ R-symmetry to be broken to a $\frak{u}(1)_R\times \frak{so}(4)_R$ subgroup. Let's see this explicitly. The $\frak{u}(1)_R$ transformations
\begin{subequations}
\begin{align}
    \hat{\calz}(\sigma, x) &= \brac{1 + i \alpha} \calz(\sigma,x) \ , \\
    \hrho_{\pm}(\sigma, x) &= \brac{1 - \frac{\alpha}{2} \Gamma_{45}} \rho_{\pm}(\sigma, x) \ , \\
    \hchi_{\pm}(\sigma, x) &= \brac{1 - \frac{\alpha}{2} \Gamma_{45}} \chi_{\pm}(\sigma, x) \ ,
\end{align}
\end{subequations}
leave the action invariant, with the conserved current
\begin{subequations}
\begin{align}
    0 &= \partial_+ J^+ + \partial_- J^- + \partial J^z + \bar{\partial} J^{\bar{z}} \ , \\
    J^{\pm} &= i \tr\brac{2 \calz D_{\mp} \bcalz - 2 D_{\mp} \calz \bcalz + \brho_{\pm} \Gamma_{45} \rho_{\pm}} \ , \\
    J^{z} &= \tr\brac{i \bar{H} \bar{Z} - 2 \brac{\bchi_- \Gamma_{0 \bar{z}} \rho_+ + \bchi_+ \Gamma_{0 \bar{z}} \rho_-}} \ .
\end{align}
\end{subequations}
The $\frak{so}(4)_R$ transformations 
\begin{subequations}
\begin{align}
    \hat{Y}^A(\sigma, x) &= \brac{Y^A + r^{AB} Y^B}(\sigma, x) \ , \\
    \hat{\rho}_{\pm}(\sigma, x) &= \brac{1 + \inv{4} r^{AB} \Gamma^{AB}} \rho_{\pm}(\sigma, x) \ , \\
    \hat{\chi}_{\pm}(\sigma, x) &= \brac{1 + \inv{4} r^{AB} \Gamma^{AB}} \chi_{\pm}(\sigma, x) \ ,
\end{align}
\end{subequations}
are also a symmetry of the action, with the associated conservation law
\begin{subequations}
\begin{align}
    0 &= \partial_+ J^{AB,+} + \partial_- J^{AB,-} + \partial J^{AB,z} + \bar{\partial} J^{AB,\bar{z}} \ , \\
    J^{AB,\pm} &= - \frac{i}{2} \tr \brac{\brho_{\pm} \Gamma^{AB} \rho_{\pm}} \ , \\
    J^{AB,z} &= \tr \brac{4 Y^A \bar{D} Y^M - i \bchi_- \Gamma^{AB} \Gamma_{0\bar{z}} \rho_+ - i \bchi_+ \Gamma^{AB} \Gamma_{0\bar{z}} \rho_- } \ .
\end{align}
\end{subequations}

As in the previous limit, the relativistic R-symmetry transformations that mix $\calz$ with $Y^A$ appear in the non-relativistic theory as a Euclidean Galilean boost in field space; the transformations
\begin{subequations}
\begin{align}
    \hat{Y}^A(\sigma, x) &= \brac{Y^A + v^A \calz + \bar{v}^A \bcalz}(\sigma, x) \ , \\
    \hat{B}(\sigma, x)  &= \brac{B - 2 [v^A \calz - \bar{v}^A \bcalz, Y^A]}(\sigma, x)  \ , \\
    \hat{H}(\sigma, x)  &= \Big( H - 8 v^A D Y^A \Big)(\sigma, x)  \ , \\
    \hchi_{\pm}(\sigma, x) &= \Big(\chi_{\pm} - \brac{\Gamma_{\calz A} \bar{v}^A + \Gamma_{\bcalz A} v^A} \rho_{\pm} \Big)(\sigma, x) \ ,
\end{align}
\end{subequations}
leave the action invariant for any holomorphic function $v^A(\sigma, z)$ with arbitrary $\sigma^{\pm}$-dependence.
The conservation law that arises from this transformation is the holomorphic condition
\begin{subequations}
\begin{align}
    0 &= \bar{\partial} j^A \ , \\
    j^A &= \tr \brac{2 \calz D Y^A - i \brho_- \Gamma_{0A z \bcalz} \rho_+} \ .
\end{align}
\end{subequations}

Finally, we note that we again have a symmetry of the theory that takes the timelike component of the gauge field into the 'large' scalar fields,
\begin{subequations}
\begin{align}
    \hat{A}_{\pm}(\sigma,x) &= \brac{A_{\pm} + \xi_{\pm} \calz + \bar{\xi}_{\pm} \bcalz}(\sigma,x) \ , \\
    \hat{B}(\sigma,x) &= \Big( B - 4i \brac{\xi_{\pm} D_{\mp} \calz - \bar{\xi}_{\pm} D_{\mp} \bcalz + \partial_{\mp} \xi_{\pm} \calz - \partial_{\mp} \bar{\xi}_{\pm} \bcalz} \Big) (\sigma,x) \ , \\
    \hat{H}(\sigma,x) &= \brac{H - 16 \xi_{\pm} F_{\mp z} }(\sigma,x) \ , \\
    \hchi_{\pm}(\sigma,x) &= \Big(\chi_{\pm} - 2 \brac{\xi_{\mp} \Gamma_{0\bcalz} + \bar{\xi}_{\mp} \Gamma_{0 \calz}} \rho_{\mp} \Big)(\sigma,x) \ .
\end{align}
\end{subequations}
However, in this case the transformation is a symmetry for any $\sigma^{\pm}$-dependent holomorphic functions $\xi_{\pm}$. As such, the conservation equation
\begin{subequations}
\begin{align}
    0 &= \bar{\partial} j_{(\pm)} \ , \\
    j_{(\pm)} &= \tr \brac{4 \calz F_{\mp z} - i \brho_{\pm} \Gamma_{z} \Gamma_{\bcalz} \rho_{\pm}} \ ,
\end{align}
\end{subequations}
leads to no conserved charge, unlike the D1NC limit discussed above.

\subsubsection{Supersymmetry} \label{sect: D3NC supersymmetry}

We now turn to the supersymmetry of the theory. It will be convenient to split the original relativistic spinor parameter into 4 components $(\alpha_{\pm}, \beta_{\pm})$, defined by
\begin{subequations}
\begin{align}
    \Gamma_{01} \alpha_{\pm} &= \pm \alpha_{\pm} \ , \\
    \Gamma_{01} \beta_{\pm} &= \pm \beta_{\pm} \ , \\
    \Gamma_{2345} \alpha_{\pm} &= \alpha_{\pm} \ , \\
    \Gamma_{2345} \beta_{\pm} &= -\beta_{\pm} \ .
\end{align}
\end{subequations}
Let us deal with $\alpha_{\pm}$ first. We can expand the spinor as
\begin{equation}
    \alpha_{\pm} = \lambda_{\pm} + \brac{\lambda_{\pm}}^* \ ,
\end{equation}
where $\lambda_{\pm}$ is defined by $i\Gamma_{23} \lambda_{\pm} = \lambda_{\pm}$. Then, the transformations
\begin{subequations}
\begin{align}
    \delta A_{\pm} &= 0 \ , \\
    \delta A_{\mp} &= i \balpha_{\pm} \rho_{\pm} \ , \\
    \delta A_z &= 0 \ , \\
    \delta \calz &= 0 \ , \\
    \delta Y^A &= - i \balpha_{\pm} \Gamma_{0A} \rho_{\mp} \ , \\ \nonumber
    \delta B &= - 4 \balpha_{\pm} \Gamma_{0\bar{z}} D \chi_{\mp} + 4 \balpha_{\pm} \Gamma_{0z} \bar{D} \chi_{\mp} - 2i \balpha_{\pm} \brac{\Gamma_{0\calz} [\calz, \chi_{\mp}] - \Gamma_{0\bcalz} [\bcalz, \chi_{\mp}]} \\
    & \hspace{0.5cm} - 4 i \partial_{\pm} \balpha_{\pm} \Gamma_{23} \rho_{\mp} \ , \\
    \delta H &= - 8i \balpha_{\pm} \Gamma_{0\calz} D \chi_{\mp} + 4 \balpha_{\pm} \Gamma_{0z} [\bcalz, \chi_{\mp}] \ , \\ \nonumber
    \delta \rho_{\pm} &= \brac{F - \inv{2} [ \calz, \bcalz] } \Gamma_{23} \alpha_{\pm} - 2 \brac{\Gamma_{\calz \bar{z}} D \calz + \Gamma_{\bcalz z} \bar{D} \bcalz} \alpha_{\pm} \\
    &\hspace{0.5cm} - 2 \brac{\Gamma_{\calz \bar{z}} \calz \partial \alpha_{\pm} + \Gamma_{\bcalz z} \bcalz \bar{\partial} \alpha_{\pm} } \ , \\
    \delta \rho_{\mp} &= 0 \ , \\
    \delta \chi_{\pm} &= \brac{-2 \bar{D} Y^A \Gamma_{Az} -2 D Y^A \Gamma_{A\bar{z}} + i [\calz, Y^A] \Gamma_{A\calz} + i[\bcalz, Y^A] \Gamma_{A \bcalz} }\alpha_{\pm} \ , \\ \nonumber
    \delta \chi_{\mp} &= - 2 \brac{2 \Gamma_{0z} F_{\pm\bar{z}} + 2 \Gamma_{0\bar{z}} F_{\pm z} + \Gamma_{0\calz} D_{\pm} \calz + \Gamma_{0\bcalz} D_{\pm} \bcalz} \alpha_{\pm} \\
    &\hspace{0.5cm} - 2 \brac{\Gamma_{0\calz} \calz + \Gamma_{0\bcalz} \bcalz} \partial_{\pm} \alpha_{\pm} \ , 
\end{align}
\end{subequations}
leave the action invariant for any $\sigma$-dependent holomorphic spinor $\lambda_{\pm} = \lambda_{\pm}(\sigma, z)$. The associated conserved current therefore takes the form
\begin{subequations}
\begin{align}
    0 &= \bar{\partial} \cal{K}_{\pm} \ , \\
    \cal{K}_{\pm} &= \tr\brac{2 F_{\pm z} \rho_{\pm} + \calz \Gamma_{0\bcalz} D \chi_{\mp} - \Gamma_{0A} \bar{D} Y^A \rho_{\mp} } \ .
\end{align}
\end{subequations}

We also need to consider the supersymmetries parameterised by $\beta_{\pm}$. The transformations
\begin{subequations}
\begin{align}
    \delta A_{\pm} &= 0 \ , \\
    \delta A_{\mp} &= i \bbeta_{\pm} \chi_{\pm} \ , \\
    \delta A_z &= - i \bbeta_{\pm} \Gamma_{0z} \rho_{\mp} \ , \\
    \delta \calz &= - 2i \bbeta_{\pm} \Gamma_{0\bcalz} \rho_{\mp} \ , \\
    \delta Y^A &= - i \bbeta_{\pm} \Gamma_{0A} \chi_{\mp} \ , \\
    \delta B &= 2i \bbeta_{\pm} \Gamma_{0A} \Gamma_{z\bar{z}} [Y^A , \chi_{\mp}] - 4 \bbeta_{\pm} \Gamma_{z\bar{z}} D_{\pm} \chi_{\pm} - 4 \partial_{\pm} \bbeta_{\pm} \Gamma_{z\bar{z}} \chi_{\pm} \ , \\
    \delta H &= -4 \bbeta_{\pm} \Gamma_{0Az\calz} [Y^A,\chi_{\mp}] - 8i \bbeta_{\pm} \Gamma_{z\calz} D_{\pm}\chi_{\pm} - 8i \partial_{\pm} \bbeta_{\pm} \Gamma_{z\calz} \chi_{\pm} \ , \\
    \delta \rho_{\pm} &= \brac{2 \Gamma_{zA} \bar{D} Y^A +2 \Gamma_{\bar{z} A} DY^A - i \Gamma_{\calz A}[\calz,Y^A] - i \Gamma_{\bcalz A} [\bcalz, Y^A]} \beta_{\pm} \ , \\
    \delta \rho_{\mp} &= -2 \brac{2\Gamma_{0z} F_{\pm\bar{z}} + 2 \Gamma_{0 \bar{z}} F_{\pm z} + \Gamma_{0\calz} D_{\pm} \calz + \Gamma_{0\bcalz} D_{\pm} \bcalz} \beta_{\pm} \ , \\
    \delta \chi_{\pm} &= \brac{ \frac{i}{2} \Gamma_{\calz \bcalz} B + \inv{2} \Gamma_{z\calz} \bar{H} + \inv{2} \Gamma_{\bar{z} \bcalz} H - \frac{i}{2} [Y^A, Y^B] \Gamma_{AB} \pm 2 F_{+-} } \beta_{\pm} \ , \\
    \delta \chi_{\mp} &= - 2 D_{\pm} Y^A \Gamma_{0A} \beta_{\pm} - 2 \Gamma_{0A} Y^A \partial_{\pm} \beta_{\pm} \ ,
\end{align}
\end{subequations}
are symmetries of the action for any $\beta_{\pm} = \beta_{\pm}(\sigma^{\pm})$. The conserved current for these symmetries are
\begin{subequations}
\begin{align}
    0 &= \partial_{\mp} \cal{S}^{\mp} + \partial \cal{S}^z + \bar{\partial} \cal{S}^{\bar{z}} \ , \\
    \cal{S}^{\mp} &= \tr\brac{2\brac{2\Gamma_{0z} F_{\pm\bar{z}} + 2 \Gamma_{0\bar{z}} F_{\pm z} + \Gamma_{0\calz} D_{\pm} \calz + \Gamma_{0\bcalz} D_{\pm} \bcalz} \rho_{\mp} }  \ , \\ \nonumber
    \cal{S}^z &= \tr \bigg( 2 \Gamma_{\bar{z} A} Y^A D_{\pm} \rho_{\pm} - 2 \Gamma_{0A} \Gamma_{z} \Gamma_{\bar{z}} \bar{D} Y^A \chi_{\mp} + i \Gamma_{0A} \Gamma_{\bcalz \bar{z}} \chi_{\mp} [\bcalz, Y^A]  \\ \nonumber
    & \hspace{0.5cm} - \frac{i}{2} \Gamma_{0\bar{z}} B \rho_{\mp} - \inv{2} \Gamma_{0\calz} \bar{H} \rho_{\mp} - \frac{i}{2} [Y^A, Y^B] \Gamma_{AB} \Gamma_{0\bar{z}} \rho_{\mp} \mp 2 F_{+-} \Gamma_{0\bar{z}} \rho_{\mp} \\
    & \hspace{0.5cm} + 4 \Gamma_z \Gamma_{\bar{z}} F_{\pm\bar{z}} \chi_{\pm}  + 2 \Gamma_{\bcalz \bar{z}} D_{\pm} \bcalz \chi_{\pm} \bigg) \ .
\end{align}
\end{subequations}

In both cases we observe an automatic superconformal enhancement of the supersymmetries, as expected from the bosonic spacetime symmetries, with the physical spacetime symmetries being those of a two-dimensional $\cal{N}=(4,4)$ SCFT.

\section{Gravitational Duals} \label{sect: gravity}

\subsection{Intersecting Brane Interpretation}

\subsubsection{The D1NC Limit} \label{sect: D1NC brane setup}

In the previous section we discussed consistent non-relativistic scaling limits of $\cal{N}=4$ SYM; in this section we look at the corresponding limits of its $AdS$ dual. We do this by first reinterpreting our field theories as arising from limits of intersecting brane set-ups before obtaining the near-horizon limits of these geometries. We will find that the solutions in both cases have the same structure as the M$p$T limits considered in \cite{Blair:2023noj}. As these limits have not been fully developed we will be somewhat schematic in this section, focusing only on expanding the relativistic solution\footnote{A discussion of the symmetries of the non-relativistic solutions, while desirable, requires an understanding of the local symmetries of the supergravity limits that we do not possess at this stage.}.

Let us first consider the D1NC limit of a D3 brane. We start with an intersecting D1-D3 geometry\footnote{We work in the string frame throughout.} with 4 relative transverse directions,
\begin{subequations}
\begin{align} \nonumber
    g &= - H_1^{-1/2} H_3^{-1/2} dt  \otimes dt  + H_1^{1/2} H_3^{-1/2} dx^i \otimes dx^i \\
    & \hspace{1.0cm} + H_1^{-1/2} H_3^{1/2} dX_s  \otimes dX_s + H_1^{1/2} H_3^{1/2} dY_s^M  \otimes dY_s^M \ , \\
    C_2 &=  H_1^{-1}   dt \wedge dX_s \ , \\
    C_4 &=  \brac{H_3^{-1} -1}   dt \wedge dx^1 \wedge dx^2 \wedge dx^3 \ , \\
    e^{\Phi} &= g_sH_1^{1/2} \ ,
\end{align}
\end{subequations}
where all fields not mentioned vanish and the functions $H_1$ and $H_3$ satisfy the equations
\begin{subequations}
\begin{align}
    0 &= \partial_M \partial_M H_1 \ , \\
    0 &= H_1 \partial_{X_s}^2 H_3 + \partial_M \partial_M H_3 \ .\label{H1eq}
\end{align}
\end{subequations}
We use the notation $X_s, Y_s^M$ for our supergravity coordinates to differentiate them from the field theory's scalar fields. Note that we are really describing a $D1$-brane smeared along the $x^i$ directions here. 

Let us first go to spatial infinity where $H_1\to h_1$ and $H_3\to h_3$ are constants. Requiring that the metric takes the form associated to a D1NC limit:
\begin{align}
g = c^2(-dt\otimes dt +dX_s\otimes dX_s)+c^{-2}(dx^i\otimes dx^i + dY_s^M\otimes dY_s^M)\ ,
\end{align}
tells us that $h_1=c^{-4}$ and $h_3=1$. Now suppose that we take $H_1$ to be completely smeared and hence 
\begin{equation}
    H_1 = c^{-4} \ ,
\end{equation}
with $c$ large. In this limit, our solution becomes
\begin{subequations} \label{eq: D1NC supergravity solution}
\begin{align} \nonumber
    g &= c^2 \brac{- H_3^{-1/2} dt  \otimes dt  + H_3^{1/2} dX_s  \otimes dX_s}\\
    & \hspace{1.0cm} + c^{-2} \brac{H_3^{-1/2} dx^i \otimes dx^i + H_3^{1/2} dY_s^M \otimes dY_s^M} \ , \\
    C_2 &= c^4 dt \wedge dX_s \ , \\
    C_4 &=  \brac{ H_3^{-1} - 1  } dt \wedge dx^1 \wedge dx^2 \wedge dx^3 \ , \\
    e^{\Phi} &= c^{-2}g_s \ ,
\end{align}
\end{subequations}
and solving (\ref{H1eq}) gives 
\begin{equation}
    H_3 = 1 + \frac{R^4}{\brac{X_s^2 + c^{-4} Y_s^M Y_s^M}^2} \ .
\end{equation}

The bosonic sector of the worldvolume theory for a single D3-brane in this geometry is described by the DBI action and brane Wess-Zumino terms,
\begin{equation}
    S_{D3} = - T_3 \int d^4 \xi e^{-\Phi} \sqrt{-\det\brac{G_{\mn} + 2\pi \alpha' F_{\mn}}} + \frac{T_3}{g_s} \int \brac{C_4 + 2\pi \alpha' C_2 \wedge F } \ .
\end{equation}
It is well-known that expanding the DBI and $C_3$ parts of this action to lowest non-trivial order and making the identifications
\begin{subequations}
\begin{align}
    X_s &= 2\pi \alpha' X \ , \\
    Y_s^M &= 2\pi \alpha' Y^M \ ,
\end{align}
\end{subequations}
between our supergravity coordinates and scalar fields gives the action of Abelian $\cal{N}=4$ SYM, where the $c$ scaling of the supergravity metric components means the field theory is defined on the spacetime with metric
\begin{equation}\label{eq: flat metric with c}
    g = - c^2 dt  \otimes dt  + c^{-2} dx^i  \otimes dx^i \ .
\end{equation}
The $C_2$ Wess-Zumino term evaluates to
\begin{equation}
    \frac{2\pi \alpha' T_3}{g_s} \int C_2 \wedge F_2 = \frac{c^4 (2\pi\alpha')^2 T_3}{2 g_s} \int dt d^3x \, \epsilon_{ijk} \partial_i X F_{jk} \ .
\end{equation}
This is a total derivative, and initially seems unimportant. However, when combined with the the DBI action we see that this term allows us to rewrite all divergent terms in a single squared quantity,
\begin{equation}
    \inv{2} F_{ij} F_{ij} - \epsilon_{ijk} \partial_i X F_{jk} + \partial_i X \partial_i X = \inv{2} \brac{F_{ij} - \epsilon_{ijk} \partial_k X}^2 \ .
\end{equation}

The non-Abelian generalisation of this is then obvious; the DBI action and the $C_4$ Wess-Zumino term give $U(N)$ $\cal{N}=4$ SYM on the scaled flat spacetime \eqref{eq: flat metric with c}, and the $C_2$ term becomes
\begin{equation}
    \frac{c^4 (2\pi\alpha')^2 T_3}{2 g_s} \tr \int dt d^3 x \, \epsilon_{ijk} D_i X F_{jk} \ ,
\end{equation}
which we note is still a total derivative and again allows us to collect all divergent terms into a single piece. Hence, defining the Yang-Mills coupling in the usual way
\begin{equation} \label{eq: YM - brane tension relation}
    g_{YM}^2 = \frac{1}{(2\pi \alpha')^2 g_s T_3}\ ,
\end{equation}
and using supersymmetry to fix the fermions we find that our leading-order action is just \eqref{eq: D1NCYM action}. Importantly, we have not had to throw away divergent boundary contributions by hand- in the brane picture there is a natural mechanism to cancel them, leading to a theory with finite-energy states.

We have seen that the leading-order terms in the $\alpha'$-expansion of the D3-brane action recovers the D1NC limit of $\cal{N}=4$ SYM. However, there are also further apparent divergences as we take $c\to\infty$ that come from higher-order terms. In order for the expansion performed here to be consistent, these must cancel once the constraint is imposed. As this is not guaranteed to be the case, the question of whether these cancellations occur is a strong test of the consistency of the D1NC limit of type IIB String Theory. We note, however, that in the the related work of \cite{Fontanella:2024rvn}, corresponding to an SNC limit,  it was argued that the $\alpha'\to 0 $ and $c\to \infty$ limits commute; this guarantees that higher-order divergences do not spoil the non-Lorentzian theory. As we will discuss below, this construction is S-dual to the D1NC limit we  consider here and we therefore also expect the higher-order derivative terms do not induce additional constraints.   We leave an exploration of these ideas to future work.

\subsubsection{The D3NC Limit} \label{sect: D3NC brane setup}

We can do the same for the D3NC limit. Consider an intersecting D3-D3' geometry with 4 relative directions,
\begin{subequations}
\begin{align} \nonumber
    g &= H_3^{-1/2} H_{3'}^{-1/2} \eta_{\alpha\beta} d \sigma^{\alpha}  \otimes d\sigma^{\beta} + H_3^{-1/2} H_{3'}^{1/2}  dx^i \otimes dx^i \\
    & \hspace{1.0cm} + H_3^{1/2}  H_{3'}^{-1/2} dX_s^a \otimes dX_s^a + H_3^{1/2} H_{3'}^{1/2} dY_s^A \otimes dY_s^A \ , \\
    C_4 &=  \brac{H_3^{-1} -1  } d\sigma^0 \wedge d\sigma^1 \wedge dx^2 \wedge dx^3\ ,\\
    C'_4 & = {H_{3'}^{-1}  } d\sigma^0 \wedge d\sigma^1 \wedge dX_s^4 \wedge dX_s^5 \ , \\
    e^{\Phi} &= g_s \ ,
\end{align}
\end{subequations}
where our indices run over the ranges $\alpha\in\{0,1\}$, $i\in\{2,3\}$, $a\in\{4,5\}$, and $A\in\{6,7,8,9\}$. We have split the contributions to the 4-form gauge fields into two pieces to isolate the contributions from the two stacks of branes.

We choose the D3'-branes to be smeared over the $x^i$ directions, so the functions $H_3$ and $H_{3'}$ satisfy the equations
\begin{subequations}
\begin{align}
    0 &= \partial_A \partial_A H_{3'} \ , \\
    0 &= H_{3'} \partial_a \partial_a H_3 + \partial_A \partial_A H_3 \ .\label{H3eq}
\end{align}
\end{subequations}
As in the previous section, we can consider a limiting case where
\begin{equation}
    H_{3'} = c^{-4}\ ,
\end{equation}
with $c$ taken to be large. Our solution then becomes
\begin{subequations} \label{eq: D3NC supergravity solution}
\begin{align} \nonumber
    g &= c^2 \brac{ H_{3}^{-1/2} \eta_{\alpha\beta} d \sigma^{\alpha} \otimes d\sigma^{\beta} + H_3^{1/2}  dX_s^a  \otimes dX_s^a } \\
    & \hspace{1.0cm}+ c^{-2} \brac{H_3^{-1/2}dx^i \otimes dx^i + H_3^{1/2}  dY_s^A \otimes dY_s^A } \ , \\
    C_4 &=  \brac{H_3^{-1} - 1  } d\sigma^0 \wedge d\sigma^1 \wedge dx^2 \wedge dx^3\ ,\\
    C'_4 & = c^4 d\sigma^0 \wedge d\sigma^1 \wedge dX_s^4 \wedge dX_s^5 \ , \\
    e^{\Phi} &= g_s \ ,
\end{align}
\end{subequations}
with the solution to (\ref{H3eq}) being
\begin{equation}
    H_3 = 1 + \frac{R^4}{\brac{X_s^a X_s^a + c^{-4} Y_s^A Y_s^A}^2} \ .
\end{equation}

Let us now look at the dynamics of the D3-brane stack. The relevant (bosonic) action for a single brane is now
\begin{equation}
    S_{D3} = - T_3 \int d^4 \xi e^{-\Phi} \sqrt{-\det\brac{G_{\mn} + 2\pi \alpha' F_{\mn}}} + \frac{T_3}{g_s} \int \brac{C_4 + C_4' } \ .
\end{equation}
The expansion of the DBI and $C_3$ terms for the solution \eqref{eq: D3NC supergravity solution} proceeds as above, with the result being that we find the bosonic action of Abelian $\cal{N}=4$ SYM on the background
\begin{equation}
    ds^2 = c^2 \eta_{\alpha\beta} d\sigma^{\alpha} d\sigma^{\beta} + c^{-2} dx^i dx^i \ .
\end{equation}
The non-Abelian generalisation of this is then just the action \eqref{eq: D3NC partial action}, with the identification \eqref{eq: YM - brane tension relation} between the gauge coupling and the brane tension. 

After pulling back $C_4'$ to the brane's worldvolume, its Wess-Zumino term is
\begin{equation}
    \frac{T_3}{g_s} \int C_4' = \frac{i c^4 \brac{2\pi \alpha'} T_3}{2 g_s} \int d^2 \sigma d^2 x \bigg( \partial_2 \brac{ \calz \partial_3 \bcalz} - \partial_3 \brac{\calz \partial_2 \bcalz} \bigg) \ ,
\end{equation}
where we have defined the complex field
\begin{equation}
    \calz = 2\pi \alpha' \brac{X_s^4 + i X_s^5} \ .
\end{equation}
When looking for the non-Abelian analogue of this term, we require that the any terms contributing to the brane's dynamics must be gauge-invariant in $C_4'$: the non-Abelian term must therefore also be a total derivative, and we find
\begin{equation}
    \frac{T_3}{g_s} \int C_4' \to \frac{i c^4 \brac{2\pi \alpha'} T_3}{2 g_s} \tr \int d^2 \sigma d^2 x \bigg( \partial_2 \brac{ \calz D_3 \bcalz} - \partial_3 \brac{\calz D_2 \bcalz} \bigg) \ .
\end{equation}
This term exactly cancels off the total derivative in \eqref{eq: D3NC total derivative bit} that one gets when rewriting the action in terms of squared quantities. We can then safely take the $c\to \infty$ limit without worrying about our states having divergent energies, leaving us with the bosonic action \eqref{eq: D3NCYM b action}. As in the D1NC limit, there are further higher-order divergences that must cancel in order for the $\alpha'$-expansion to be consistent; we again leave this to future work.

\subsection{Near-Horizon Geometries}

\subsubsection{The D1NC Limit}

We have seen that the intersecting brane set-ups considered above reproduce the non-relativistic field theories discussed in section \ref{sect: field theory}. Let us now consider the supergravity solutions that arise from these. To simplify our notation we will drop the subscript on the supergravity coordinates from here onwards and use hats to denote any variable that contains $c$.

As seen in \eqref{eq: D1NC supergravity solution}, the metric in the large $c$ limit has the decomposition
\begin{subequations} \label{eq: metric c expansion}
\begin{align}
    \hat{g} &= c^2 \hat{\tau}_{\mn} dx^{\mu} \otimes dx^{\nu} + c^{-2} \hat{h}_{\mn} dx^{\mu} \otimes dx^{\nu} \ , \\
    \hat{\tau}_{\mn} dx^{\mu} \otimes dx^{\nu} &= -\hat{H}^{-1/2} dt \otimes dt + \hat{H}^{1/2} d X \otimes dX \ , \\
    \hat{h}_{\mn} dx^{\mu} \otimes dx^{\nu} &= H^{-1/2} dx^i \otimes dx^i + \hat{H}^{1/2} dY^A \otimes dY^A \ , \\
    \hat{H} &= 1 + \frac{R^4}{\brac{X^2 + c^{-4} Y^M Y^M}^2 }  \ , 
\end{align}
\end{subequations}
with the corresponding form
\begin{subequations} \label{eq: cometric c expansion}
\begin{align}
    \hat{g}^{-1} &= c^2 \hat{h}^{\mn} \partial_{\mu} \otimes \partial_{\nu} + c^{-2} \hat{\tau}^{\mn} \partial_{\mu} \otimes \partial_{\nu} \ , \\
    \hat{h}^{\mn} \partial_{\mu} \otimes \partial_{\nu} &= \hat{H}^{1/2} \partial_i \otimes \partial_i + \hat{H}^{-1/2} \partial_M \otimes \partial_M \ , \\
    \hat{\tau}^{\mn} \partial_{\mu} \otimes \partial_{\nu} &= - \hat{H}^{1/2} \partial_t \otimes \partial_t + \hat{H}^{-1/2} \partial_X \otimes \partial_X \ ,
\end{align}
\end{subequations}
for the inverse metric. The relativistic metric has split into $p$-brane Newton-Cartan fields \cite{Bergshoeff:2023rkk}; when the $c\to\infty$ limit is performed the well-defined leading order tensor fields  arise from $\hat{\tau}_{\mn}$ and $\hat{h}^{\mn}$, so we shall focus on this index configuration. We see that $\hat{\tau}_{\mn}$ is a Lorentzian 2-metric along the D1-brane's longitudinal directions, while $\hat{h}^{\mn}$ is a Riemannian 8-cometric in the transverse directions. The expansion of the metric fields in powers of $c^{-4}$ is
\begin{subequations}
\begin{align}
    \hat{\tau}_{\mn} dx^{\mu} \otimes dx^{\nu} &= \tau_{\mn} dx^{\mu} \otimes dx^{\nu} + c^{-4} \eta_{mn} \brac{\tau^{m} \otimes m^{n} + m^{m} \otimes \tau^{n}} + O(c^{-8}) \ , \\
    \hat{h}^{\mn} \partial_{\mu} \otimes \partial_{\nu} &= h^{\mn} \partial_{\mu} \otimes \partial_{\nu} + c^{-4} \delta^{IJ} \brac{e_{I} \otimes \pi_{J} + \pi_{J} \otimes e_I } + O(c^{-8}) \ ,
\end{align}
\end{subequations}
where $\{\tau^m\}$ and $\{e_I\}$ are vielbeins for $\tau_{\mn}$ and $h^{\mn}$ respectively.

We now take the near-horizon limit, where
\begin{equation}
    \hat{H} \to \frac{R^4}{\brac{X^2 + c^{-4} Y^M Y^M}^2} \ ,
\end{equation}
The limits of \eqref{eq: metric c expansion} and \eqref{eq: cometric c expansion} give the Newton-Cartan metric structures
\begin{subequations}
\begin{align}
    \tau_{\mn} dx^{\mu} \otimes dx^{\nu} &= - \frac{X^2 }{R^2} dt\otimes dt + \frac{R^2}{X^2}  dX \otimes dX \ , \\
    h^{\mn} \partial_{\mu} \otimes \partial_{\nu} &= \frac{R^2}{X^2} \partial_i \otimes \partial_i + \frac{X^2}{R^2 } \partial_M \otimes \partial_M \ .
\end{align}
\end{subequations}
We recognise the geometry given by the Lorentzian metric as $AdS_2$, while $h$ defines a pair of planes with overall scale factors that grow and shrink as $X$, the $AdS_2$ radial coordinate, varies. If we choose vielbeins
\begin{subequations}
\begin{align}
    \tau^t &= \frac{X}{R} dt \ , \\
    \tau^X &= \frac{R}{X} dX \ , \\
    e_i &= \frac{R}{X} \partial_i \ , \\
    e_M &= \frac{X}{R} \partial_M \ ,
\end{align}
\end{subequations}
for these tensors, the subleading metric fields take the form
\begin{subequations}
\begin{align}
    m^t &= \frac{ Y^A Y^A}{2 R X} dt \ , \\
    m^X &= -\frac{R Y^A Y^A}{2 X^3} dX \ , \\
    \pi_i &= - \frac{R Y^A Y^A}{2 X^3} \partial_i \ , \\
    \pi_M &= \frac{Y^B Y^B}{2 R X} \partial_M \ .
\end{align}
\end{subequations}

The 5-form field strength is
\begin{subequations}
\begin{align}
    \hat{F}_5 &= \brac{1 + \star} d\hat{C}_4 \ , \\
    \hat{C}_4 &=  \brac{\hat{H}^{-1} -1} dt \wedge dx^1 \wedge dx^2 \wedge dx^3 \ ,
\end{align}
\end{subequations}
which explicitly evaluates to
\begin{align} \nonumber
    \hat{F}_5 = \frac{4 R^4}{\brac{X^2 + c^{-4} Y^A Y^A}^3} \bigg[& H^{-2} dt \wedge dx^1 \wedge dx^2 \wedge dx^3 \wedge \big(X dX  + c^{-4}  Y^M d Y^M \big) \\ \nonumber
    &+ c^{-4} \sum_{M=5}^9 (-1)^M Y^M dX \wedge d Y^5 \wedge ... \wedge d \check{Y}^M \wedge ... \wedge dY^9 \\
    &+ c^{-4} X dY^5 \wedge ... \wedge d Y^9
    \bigg] \ ,
\end{align}
where we use $d\check{Y}^M$ to denote the omission of $dY^M$ from the product. Hence, we can take the near-horizon limit and introduce the expansion
\begin{equation}
    \hat{F}_5 = F_5 + c^{-4} \Tilde{F}_5 + O\brac{c^{-8}}
\end{equation}
to get
\begin{subequations}
\begin{align}
     F_5 &= \frac{4 X^3}{R^4} dt \wedge dx^1 \wedge dx^2 \wedge dx^3 \wedge dX \ , \\ \nonumber
     \Tilde{F}_5 &= \frac{4}{R^4} dt \wedge dx^1 \wedge dx^2 \wedge dx^3 \wedge \brac{Y^M Y^M X dX + X^2 Y^M dY^M} \\ \nonumber
     &\hspace{0.5cm} + \frac{4R^4}{X^6} \bigg(\sum_{M=5}^9 (-1)^M Y^M dX \wedge d Y^5 \wedge ... \wedge d \check{Y}^M \wedge ... \wedge dY^9 \\
     &\hspace{0.5cm}+ X dY^5 \wedge ... \wedge d Y^9 \bigg) \ .
\end{align}
\end{subequations}
We note in passing that if we started with a different index configuration for the relativistic field the relative weightings with $c$ of the terms would differ from that observed here.

The last two non-trivial fields in the supergravity solution \eqref{eq: D1NC supergravity solution} are the constant diverging $C_2$ field and dilaton, from which we extract the $c$-dependence by writing it in the form $e^{\hat \phi} = c^{-2} g_s e^{\varphi}$ with $\varphi=0$. We note that we can therefore write $C_2$ as
\begin{equation}
    C_2 = c^4 e^{-\phi} \tau^t \wedge \tau^X \ ,
\end{equation}
which is the required form for the M$1$T limit found by S-dualising the SNC limit \cite{Blair:2023noj}.

\subsubsection{The D3NC Limit}

The same analysis can be done for the D3NC limit using the supergravity solution \eqref{eq: D3NC supergravity solution}. Rewriting $\{X^4, X^5\}$ as
\begin{subequations}
\begin{align}
    X^4 &= r \cos\theta \ , \\
    X^5 &= r \sin\theta \ , 
\end{align}
\end{subequations}
for convenience, taking both the near-horizon and $c\to\infty$ limits gives the Newton-Cartan metric structures
\begin{subequations}
\begin{align}
    \tau_{\mn} dx^{\mu} \otimes dx^{\nu} &=  \frac{r^2}{R^2} \eta_{\alpha\beta} d\sigma^{\alpha} \otimes d\sigma^{\beta} + \frac{R^2}{r^2} dr \otimes dr + R^2 d\theta \otimes d\theta \ , \\
    h^{\mn} \partial_{\mu} \otimes \partial_{\nu} &= \frac{R^2}{r^2} \partial_i \otimes \partial_i + \frac{r^2}{R^2} \partial_A \otimes \partial_A \ ,
\end{align}
\end{subequations}
where $\tau_{\mn}$ is a Lorentzian 4-metric and $h^{\mn}$ a Riemannian 6-cometric. Using the vielbeins
\begin{subequations}
\begin{align}
    \tau^{\alpha} &= \frac{r}{R} d\sigma^{\alpha} \ , \\
    \tau^r &= \frac{R}{r} dr \ , \\
    \tau^{\theta} &= R d\theta \ , \\
    e_i &= \frac{R}{r} \partial_i \ , \\
    e_A &= \frac{r}{R} \partial_A \ ,
\end{align}
\end{subequations}
the subleading metric fields are
\begin{subequations}
\begin{align}
    m^{\alpha} &= \frac{ Y^A Y^A}{2 R r} d\sigma^{\alpha} \ , \\
    m^r &= -\frac{R Y^A Y^A}{2 r^3} dr \ , \\
    m^{\theta} &= - \frac{R Y^A Y^A}{2 r^2} d\theta \ , \\ 
    \pi_i &= - \frac{R Y^A Y^A}{2 r^3} \partial_i \ , \\
    \pi_A &= \frac{Y^B Y^B}{2 R r} \partial_A \ .
\end{align}
\end{subequations}
The geometry defined by $\tau$ in the near-horizon limit is $AdS_3\times S^1$, and as before the geometry defined by $h$ consists of two planes that grow and shrink as the $AdS_3$ radial coordinate $r$ varies.

The 5-form field strength in the near-horizon limit has the leading and subleading components
\begin{subequations}
\begin{align}
    F_5 &= \frac{4 r^3}{R^4} d\sigma^0 \wedge ... \wedge dx^3 \wedge dr \ , \\ \nonumber
    \Tilde{F}_5 &= \frac{4}{R^4} d\sigma^0 \wedge d\sigma^1 \wedge dx^2 \wedge dx^3 \wedge \brac{Y^A Y^A r dr + r^2 Y^A dY^A}\\ \nonumber
    &\hspace{0.5cm} + \frac{4 R^4}{ r^6} \bigg(\sum_{M=6}^9 (-1)^M Y^M dX^4 \wedge dX^5 \wedge dY^6 \wedge ... \wedge d\check{Y}^M \wedge ... \wedge dY^9 \\
    &\hspace{2cm}+ r^2 d\theta \wedge dY^6 \wedge ... \wedge dY^9  \bigg) \ .
\end{align}
\end{subequations}
There is also the constant diverging field $C_4'$ that does not contribute to $F_5$. However, similarly to the D1NC case, we note that it can be written as
\begin{equation}
    C_4' = c^4 e^{-\varphi} \tau^0 \wedge \tau^1 \wedge \tau^r \wedge \tau^{\theta} \ ,
\end{equation}
where $\varphi=0$ is the dilaton of the solution excluding the string coupling, exactly as required for the M$3$T limit of \cite{Blair:2023noj}.

\section{Relating Theories} \label{sect: relations}

\subsection{Dimensional Reduction and T-Duality}\label{sec: T-duality}

In section \ref{sect: field theory} described two non-trivial non-relativistic limits of $\cal{N}=4$ super-Yang-Mills and we may wonder if they are related in some way. Here we will show that this is the case: dimensionally reducing the D1NC theory on a 'small' spatial direction and the D3NC theory on the 'large' spatial direction leads to the same three-dimensional theory.

We first review the dimensional reduction of the D1NC action \eqref{eq: D1NCYM action}, which was performed in \cite{Lambert:2018lgt}. Reducing along the $x^3$ direction and using the notation
\begin{subequations}
\begin{align}
    x^i &= (x^{\alpha}, x^3) \\
    X &\equiv X_1 \ , \\
    A_3 &\equiv X_2 \ , \\
    F_{12} & \equiv F \ ,
\end{align}
\end{subequations}
gives
\begin{align} \nonumber
    S_{D1NC,R} = \frac{2\pi R_3}{g_{D1}^2} \tr \int dt d^2x \bigg(&
    F_{0\alpha} F_{0\alpha} + \brac{D_t X_1}^2 + \brac{D_t X_2}^2 + 2 G_{12} \brac{F - i [X_1, X_2]} \\ \nonumber
    &+ 2 G_{\alpha 3} \brac{D_{\alpha} X_2 + \epsilon_{\alpha\beta} D_{\beta} X_1} - D_{\alpha} Y^M D_{\alpha} Y^M \\ \nonumber
    &+ [X_1,Y^M]^2 + [X_2, Y^M]^2 - i \bpsi_+ D_t \psi_+ - 2i \bpsi_- \Gamma_{0\alpha} D_{\alpha} \psi_+ \\ 
    &- 2 \bpsi_- \Gamma_{03} [X_2, \psi_+] - 2 \bpsi_- \Gamma_{04} [X_1, \psi_+] + \bpsi_+ \Gamma^M [Y^M, \psi_+]
    \bigg) \ .
\end{align}
Working in complex coordinates $z = x^1 + i x^2$ in the spatial directions and defining the fields
\begin{subequations}
\begin{align}
    \calz &= X^1 + i X^2  \ , \\
    B &= - 2 G_{12} \ , \\
    H &= i G_{13} + G_{23} \ , 
\end{align}
\end{subequations}
this becomes
\begin{align} \nonumber
    S_{D1NC,R} = \frac{2\pi R_3}{g_{D1}^2} \tr \int dt d^2x \bigg(&
    4 F_{0z} F_{0\bar{z}} + D_t \calz D_t \bcalz - B \brac{F + \inv{2} [\calz, \bcalz] }  - H \bar{D} \calz \\ \nonumber
    &- \bar{H} D \bcalz - 4 D Y^M \bar{D} Y^M + [\calz, Y^M] [ \bcalz, Y^M] - i \bpsi_+ D_t \psi_+ \\ \nonumber
    &- 2i \bpsi_- \brac{\Gamma_{01} + i \Gamma_{02}}D \psi_+ - 2i \bpsi_- \brac{\Gamma_{01} - i \Gamma_{02}}\bar{D} \psi_+ \\ \nonumber
    &-  \bpsi_- \brac{ \Gamma_{04} - i \Gamma_{03}} [\calz, \psi_+] -  \bpsi_- \brac{ \Gamma_{04} + i \Gamma_{03}} [\bcalz, \psi_+] \\ 
    &+ \bpsi_+ \Gamma^M [Y^M, \psi_+]
    \bigg) \ .
\end{align}

Let us now dimensionally reduce the D3NC theory. We will reduce along the $\sigma^1$ direction, so it will be convenient to undo the split of the fermions into the eigenspaces of $\Gamma_{01}$. Defining
\begin{subequations}
\begin{align}
    A_1 &\equiv Y^5 \ , \\
    Y^M &= (Y^A, Y^5) \ ,
\end{align}
\end{subequations}
gives
\begin{subequations}
\begin{align} \nonumber
    S_{D3NC,R} = \frac{2\pi R_1}{g_{D3}^2} \tr \int dt d^2x \bigg(&
    4 F_{0z} F_{0\bar{z}} + D_t \calz D_t \bcalz - B \brac{F + \inv{2} [\calz, \bcalz]} - H \bar{D} \calz \\ \nonumber 
    &- \bar{H} D \bcalz - 4 D Y^M \bar{D} Y^M + [\calz, Y^M] [\bcalz, Y^M] - i \brho D_t \rho \\ \nonumber
    &- 4 i \bchi \brac{\Gamma_{0\bar{z}} D + \Gamma_{0z} \bar{D}} \rho -2 \bchi \Gamma_{0\calz} [\calz, \rho] - 2\bchi \Gamma_{0 \bcalz} [\bcalz, \rho] \\
    &- \brho \Gamma_{01} [Y^5, \rho] - \brho \Gamma_{0A} [Y^A, \rho]
    \bigg) \ .
\end{align}
\end{subequations}
We immediately see that the bosonic terms in both actions match if we take the couplings to satisfy the relation
\begin{equation}
    \frac{g_{D1}^2}{R_3} = \frac{g_{D3}^2}{R_1} \ .
\end{equation}
Redefining our D1NC fermions with the transformation
\begin{subequations}
\begin{align}
    \psi_+ &= \inv{\sqrt{2}}\brac{\bbm{1} + \Gamma_{01234}} \rho \ , \\
    \psi_- &= \inv{\sqrt{2}}\brac{\bbm{1} + \Gamma_{01234}} \chi \ ,
\end{align}
\end{subequations}
and using the gamma matrix combinations
\begin{subequations}
\begin{align}
    \Gamma_{0z}^{(D1NC)} &= \inv{2} \brac{\Gamma_{01} - i \Gamma_{02}} \ , \\
    \Gamma_{0\calz}^{(D1NC)} &= \inv{2} \brac{\Gamma_{04} - i \Gamma_{03}} \ ,
\end{align}
\end{subequations}
it is also clear that the fermionic terms are also identical and the two theories are equal after dimensional reduction.

There is a natural interpretation of this in terms of T-duality. Our string-theoretic picture is that the theories arise from considering intersections of D3-branes with DpNC branes. In the D1NC limit this comes from the brane setup
\begin{align} \label{eq: D1NC brane setup}
\begin{array}{rrrrrr}
D3: & 0 & 1 &2&3& \\
D1NC:& 0 &  & & &4 \ \ , \\
\end{array}
\end{align}
and in the D3NC limit we're considering
\begin{align} \label{eq: D3NC brane setup}
\begin{array}{rrrrrrr}
D3: & 0 & 1 &2& 3 && \\
D3NC:& 0 & 1 & & &4&5 \ \ .
\end{array}
\end{align}
Suppose we T-dualise the D1NC setup along the $x^3$ direction; this is a longitudinal direction for the D3-brane and transverse for the D1NC-brane, so the usual rules of T-duality convert the brane diagram \eqref{eq: D1NC brane setup} into
\begin{align} \label{eq: D2NC brane setup}
\begin{array}{rrrrrr}
D2: & 0 & 1 &2& & \\
D2NC:& 0 &  & &3 &4 \ \ . \\
\end{array}
\end{align}
However, if we T-dualise the D3NC setup along the $x^1$ direction then, as this direction is longitudinal for both branes, we find that \eqref{eq: D3NC brane setup} becomes
\begin{align}
\begin{array}{rrrrrrr}
D2: & 0 &  &2&3 & & \\
D2NC:& 0 &  & & &4 & 5 \ \ . \\
\end{array}
\end{align}
Upon relabelling the coordinates, we see that this is identical to \eqref{eq: D2NC brane setup}; the limits are therefore T-dual to each other.

More generally we see that we can perform T-dualities in a variety of directions. In particular T-duality along a worlvolume direction of a D$p$-brane is simply dimensional reduction. The resulting theories will be dynamical in the large directions, {\it i.e.} along the intersection of the D$p$-brane and D$q$NC probe brane, and the dynamics will localise onto the moduli space of BPS solutions in the remaining small directions. The results of some T-dualites are given in figure 1. In particular the additional theories on D2-branes and D3-branes, as well as the D0NC limit of D4-branes, were explicitly constructed in \cite{Lambert:2018lgt}.

 \begin{figure}
    \centering
    \includegraphics[width=1\linewidth]{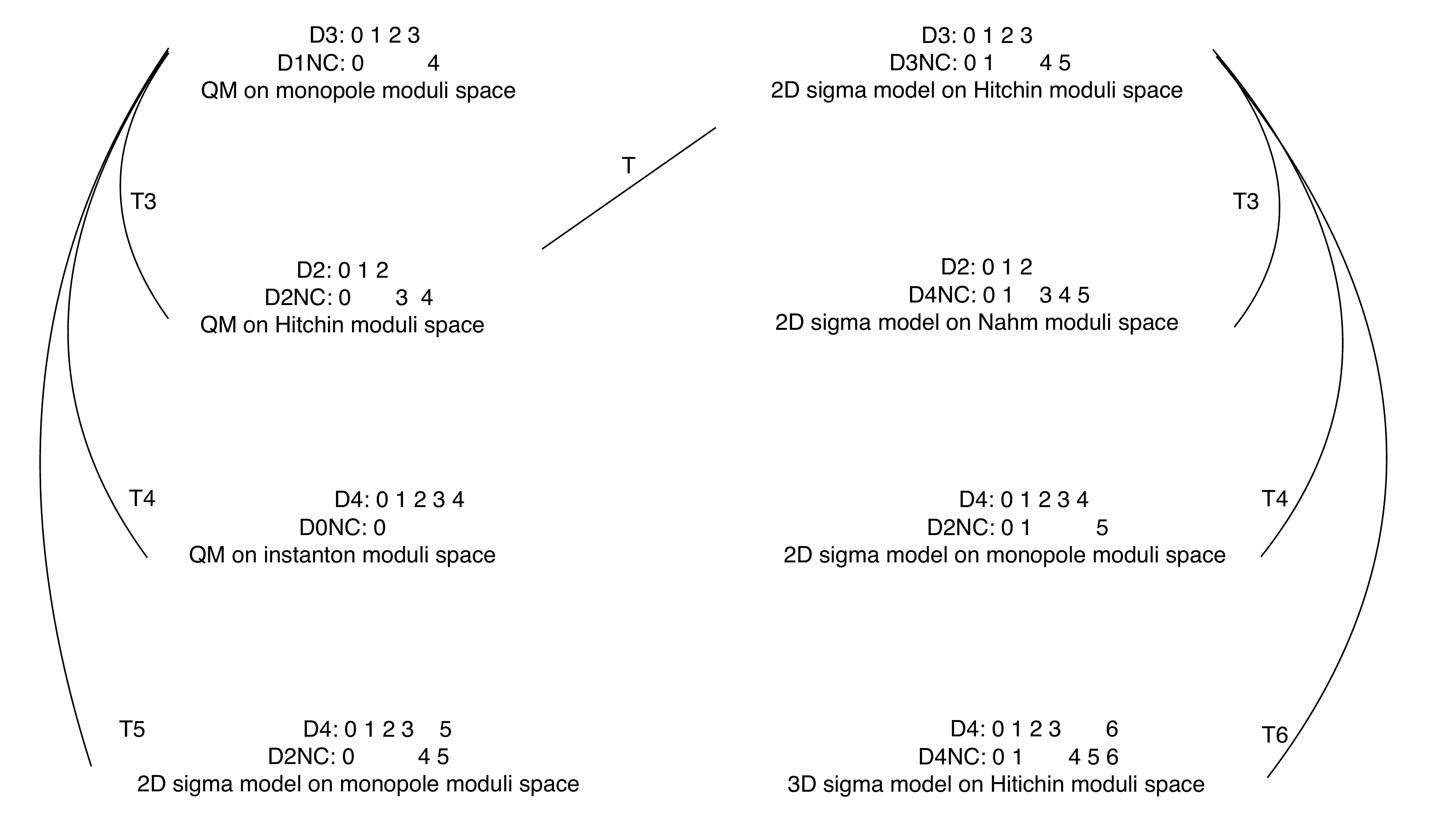}
    \caption{T-duality Web: The duality between the two sides is explicitly given in section \ref{sec: T-duality} and the two 2D sigma model examples are equivalent as a consequence of T-duality.}
    \label{fig:enter-label}
\end{figure}

\subsection{D1NC and Supersymmetric Galilean Yang-Mills} \label{sect: D1NC and SGYM}

It is well-known that one can obtain non-relativistic field theories from the null reductions of Lorentzian theories \cite{Baiguera:2022cbp, Baiguera:2023fus}. Let us consider five-dimensional $\cal{N}=2$ SYM in lightcone coordinates,
\begin{align} \nonumber
    S_{5d} = \inv{2g_{5d}^2} \tr \int dx^+ dx^- d^3 x \bigg[& F_{+-}^2 +2 F_{+i} F_{-i} - \inv{2} F_{ij} F_{ij} + 2 D_+ Y^M D_- Y^M \\ \nonumber
    &- D_i Y^M D_i Y^M + \inv{2} [Y^M, Y^N]^2 - \sqrt{2} i \bpsi \Gamma_0 \Gamma_- D_+ \psi \\ \label{eq: five-dimensional N=2 action}
    &- \sqrt{2} i \bpsi \Gamma_0 \Gamma_+ D_- \psi - i \bpsi \Gamma_0 \Gamma_i D_i \psi + \bpsi \Gamma_0  \Gamma^M [Y^M, \psi] \bigg] \ .
\end{align}
For definiteness we use
\begin{equation}
    x^{\pm} = \inv{\sqrt{2}} \brac{x^0 \pm x^4}\ ,
\end{equation}
for our lightcone coordinates, so
\begin{equation}
    \Gamma_{\pm} = \inv{2} \brac{\Gamma_{0} \pm \Gamma_4} \ .
\end{equation}
where we use the same spinor conventions as in \eqref{eq: rel N=4 action 2}. Suppose we reduce this action on the null coordinate $x^+$, keeping only the zero-modes; using the notation $A_+ \equiv X$ for the component of the gauge field along this direction, relabelling $x^-$ to $t$, and defining
\begin{equation}
    \psi_{\pm} = \inv{2} \brac{\bbm{1} \mp \Gamma_{04}} \psi \ ,
\end{equation}
we get
\begin{align} \nonumber
    S_{SGYM} = \frac{\pi R_+}{g_{5d}^2} \tr \int dt d^3 x \bigg[& D_0 X D_0 X - 2 D_i X F_{0i} - \inv{2} F_{ij} F_{ij} - 2i D_0 Y^M [X,Y^M] \\ \nonumber
    &- D_i Y^M D_i Y^M + \inv{2} [Y^M, Y^N]^2 + \sqrt{2} i \bpsi_+  D_0 \psi_+ \\ \label{eq: SGYM action}
    &-  2i \bpsi_- \Gamma_{0i} D_i \psi_+ + \sqrt{2}  \bpsi_-  [X,\psi_-] + 2 \bpsi_- \Gamma_{0M}  
    [Y^M , \psi_+] \bigg] \ .
\end{align}
The bosonic sector of the action is four-dimensional Galilean Yang-Mills, which was first studied in \cite{Bagchi:2015qcw} and obtained via null reduction in \cite{Bagchi:2022twx}, coupled to adjoint-valued scalar fields $Y^M$ in the fundamental of $SO(5)$; interestingly, this found to arise by taking an SNC limit\footnote{{\it i.e.} a non-relativistic limit with a critical $B$-field.} of the D3-brane's non-Abelian DBI action \cite{Fontanella:2024rvn}. As the SNC and D1NC limits are related by S-duality one may wonder whether there is a relation between the D1NC limit discussed in section \ref{sect: D1NC} and this theory.

As a first point of comparison, we should examine the symmetries of \eqref{eq: SGYM action} and compare them to those found in section \ref{sect: D1NC symmetries}. In \cite{Bagchi:2022twx} it was found that the spacetime symmetries of the pure Galilean Yang-Mills action are identical to those of the D1NC theory: the 'physical' transformations (those with non-vanishing Noether charges) are the $SO(2,1)$ transformations \eqref{eq: D1NC time transformations} and $SO(3)$ rotations \eqref{eq: D1NC rotations}, while the time-dependent spatial translations \eqref{eq: D1NC spatial translations} are 'unphysical'. Let us extend this to the full action. A short calculation shows the $SO(2,1)$ transformations are symmetries provided we take the field transformations
\begin{subequations}
\begin{align}
    \hat{X}(\hat{t},\hat{x}) &= \brac{1 - \df} X(t,x) \ , \\
    \hat{Y}^M(\hat{t},\hat{x}) &= \brac{1 - \df} Y^M(t,x) \ , \\
    \hat{A}_t(\hat{t}, \hat{x}) &= \brac{\brac{1 - \df} A_t - \ddf x^i A_i}(t,x) \ , \\
    \hat{A}_i(\hat{t}, \hat{x}) &= \brac{\brac{1 - \df} A_i - \ddf x^i X}(t,x) \ , \\
    \hpsi_+(\hat{t}, \hat{x}) &= \brac{1 - \frac{3}{2} \df} \psi_+(t,x) \ , \\
    \hpsi_-(\hat{t}, \hat{x}) &= \brac{\brac{1 - \frac{3}{2}\df} - \inv{\sqrt{2}} \ddf \Gamma_{0i} x^i \psi_+}(t,x) \ .
\end{align}
\end{subequations}
Similarly, the time-dependent translations are symmetries for the field transformations
\begin{subequations}
\begin{align}
    \hat{X}(\hat{t},\hat{x}) &= X(t,x) \ , \\
    \hat{Y}^M(\hat{t},\hat{x}) &=  Y^M(t,x) \ , \\
    \hat{A}_t(\hat{t}, \hat{x}) &= \brac{ A_t - \dot{\xi}^i  A_i}(t,x) \ , \\
    \hat{A}_i(\hat{t}, \hat{x}) &= \brac{ A_i - \dot{\xi}^i X}(t,x) \ , \\
    \hpsi_+(\hat{t}, \hat{x}) &= \psi_+(t,x) \ , \\
    \hpsi_-(\hat{t}, \hat{x}) &= \brac{\psi_- - \inv{\sqrt{2}}  \Gamma_{0i} \dot{\xi}^i \psi_+}(t,x) \ .
\end{align}
\end{subequations}
The final set of spacetime transformations, rotations, are symmetries when the fields transform as in \eqref{eq: rotations field transformations}. The action also has an $SO(5)$ R-symmetry with the same field transformations as \eqref{eq: SO(5) R symmetry}. 

Aside from the exotic symmetry \eqref{eq: strange transformation} there is an obvious matching of the physical bosonic symmetries between the D1NC theory and the SGYM theory, hinting at a deeper relation between the two. This comes from interpreting both theories as reductions of the six-dimensional (2,0) theory. Let us start with the six-dimensional theory on the DLCQ background 
\begin{equation}
    ds^2 = - 2 dx^+ dx^- + dx^i dx^i + R^2 d\theta^2 \ ,
\end{equation}
where we periodically identify $x^+\sim x^+ + 2\pi R_+$ and $\theta \sim \theta + 2\pi$. The theory in this regime is known \cite{Aharony:1997an, Aharony:1997th} to be quantum mechanics on the moduli space of instantons on $\bb{R}^3 \times S_R^1$. This can also be written as a five-dimensional non-Lorentzian gauge theory \cite{Lambert:2018lgt}  for which $g_5^2 = 4\pi^2 R_+$. If we take $R\to 0$ to reduce on the compact direction whilst also keeping the ratio $R_+/R = k$ finite we recover the D1NC theory with coupling $g_{D1}^2 = 4\pi k$. However, suppose we instead started by taking the $R\to 0$ limit: we would then have weakly-coupled five-dimensional $\cal{N}=2$ SYM on a flat background with a periodic null direction. If we take $R_+\to 0$ then we perform a null reduction, with the zero-modes giving \eqref{eq: SGYM action} (for a discussion of the higher Fourier modes see appendix \ref{sect: YM null reduction}). If we are more careful and again take the limit with $R_+/R = k$ finite we see that the coupling is $g_{SGYM}^2 = \frac{4\pi}{k}$. As the order of compactification is irrelevant, the two theories should be dual to one another. The couplings of the theories are (up to a constant) inverses of each other, as expected from an S-duality transformation. It would be interesting to test this, for example by computing the supersymmetry of \eqref{eq: SGYM action} and comparing it to the D1NC results.

\section{Conclusion} \label{sect: conclusions}

In this paper we have analysed  two non-relativistic limits, which we referred  to as D1NC and D3NC, of four-dimensional ${{\cal N}=4}$ super-Yang-Mills. We interpreted these limits  as arising from intersections of a stack of D3-branes with a non-relativistic probe D1-branes or D3-branes respectively. We saw that the resulting field theories have an infinite dimensional symmetry group and that their dynamics subsequently leads to Quantum Mechanics on monopole moduli space or a two-dimensional sigma-model on Hitchin moduli space. We also considered the corresponding limits of the dual AdS geometries which are described by Newton-Cartan limits of   type IIB supergravity.

The field theories constructed here, and also in \cite{Lambert:2024uue}, have intriguing local symmetries which we expect should be treated as gauge symmetries. In particular this means that the only physical states are invariant under the local symmetries. Furthermore we expect that only the rigid symmetries need to match with the symmetries of the AdS dual. We have seen that half of the supersymmetries are local and this would suggest that they don't need to be visible in the gravity dual. Thus it could well be that the supersymmetric completion of Newton-Cartan type IIB supergravity \cite{Bergshoeff:2023ogz} (and also eleven-dimensional supergravity in the case of M2-branes \cite{Blair:2021waq}) may only have half the maximal amount of supersymmetry, {\it i.e.} sixteen supercharges. In future work we hope to greater explore the manifestation of symmetries in the AdS  duals and test whether or not the AdS/CFT correspondence survives these non-relativistic limits.

Lastly, we would like to comment on the relation of our work to the recent paper \cite{Fontanella:2024rvn} which explores an SNC limit of D3-branes. This is S-dual to the D1NC limit we considered here. As discussed above this suggests that the Galilean Super-Yang-Mills theory they obtained is related by an S-duality to the non-relativistic theory we constructed from the D1NC limit. It is curious to note that Galilean Super-Yang-Mills does not have any constraints beyond the Gauss law whereas the theory we constructed has a constraint that restricts the dynamics to the moduli space of BPS monopoles. We hope to address whether or not S-duality relates these theories in greater detail in future work.

\section*{Acknowledgments}

We would like to thank E. Bergshoeff and C. Blair for informative discussions. N.L. is supported in part  by the STFC consolidated grant ST/X000753/1. J.S. is supported by the STFC studentship ST/W507556/1.

\appendix

\section{Null Reduction of super-Yang-Mills} \label{sect: YM null reduction}

In section \ref{sect: D1NC and SGYM} we discussed a proposed duality between the D1NC theory and the null reduction of five-dimensional $\cal{N}=2$ SYM. Here we will compute the bosonic part of the reduction with all Kaluza-Klein modes retained, where we expand the five-dimensional fields in the Fourier modes
\begin{subequations}
\begin{align}
    A_+(x^+,x_{4d}) &= \sum_{n} e^{- \frac{i n x^+}{R_+}} X^{(n)}(x_{4d}) \ , \\
    A_-(x^+, x_{4d}) &= \sum_{n} e^{- \frac{i n x^+}{R_+}} A_-^{(n)}(x_{4d}) \ , \\
    A_i(x^+, x_{4d}) &= \sum_{n} e^{- \frac{i n x^+}{R_+}} A_i^{(n)}(x_{4d}) \ , \\
    Y^M(x^+, x_{4d}) &= \sum_{n} e^{- \frac{i n x^+}{R_+}} Y^M_{(n)}(x_{4d}) \ .
\end{align}
\end{subequations}
Our task is to plug this expansion into \eqref{eq: five-dimensional N=2 action} and take the $R_+\to 0$ limit.

Starting with the $F_{+-}^2$ term, integrating over the compact coordinate gives
\begin{align} \nonumber
    \tr \int \frac{dx^+}{2\pi R_+} \, F_{+-}^2 = \tr \bigg( &
    \sum_n \abs{ \frac{i n}{R_+} A_-^{(n)} + \partial_- X^{(n)} }^2  - \sum_{n,m,p} [A_-^{(n)}, X^{(m)}] [A_-^{(p)}, \bar{X}^{(n+m+p)}] \\ \nonumber
    &+ 2i \sum_{n,m} \brac{\frac{in}{R_+} A_-^{(n)}  + \partial_- X^{(n)} } [\bar{X}^{(m+n)}, A_-^{(m)}]
    \bigg) \ .
\end{align}
The $A_-^{(n)}$ tower of fields acquire the standard Kaluza-Klein masses
\begin{equation}
    m^2_n = \frac{n^2}{R_+^2} \ ,
\end{equation}
so taking $R_+\to 0$ localises the path integral onto configurations for which
\begin{equation}
    A_-^{(n)} = 0 \ \forall \ n\neq0 \ .
\end{equation}
We'll assume this from here onwards, and relabel $A_-^{(0)}$ to $A_-$ for convenience. Hence, in the $R_+\to 0$ limit we have
\begin{equation}
    \tr \int \frac{dx^+}{2\pi R_+} \, F_{+-}^2 \to \cal{L}_1 =  \tr \sum_n D_- X^{(n)} D_- \bar{X}^{(n)} \ .
\end{equation}
Doing the same for the $F_{+i} F_{-i}$ term gives
\begin{align} \nonumber
    \tr \int \frac{dx^+}{2\pi R_+} \, F_{+i} F_{-i} = \tr \bigg(& 
    \partial_i A_- \partial_i X^{(0)}  - \sum_n D_- A_i^{(n)} \brac{\frac{i n}{R_+} \bar{A}_i^{(n)} + \partial_i \bar{X}^{(n)}} \\
    &- i \partial_i A_- \sum_n [A_i^{(n)}, \bar{X}^{(n)}] + i \sum_{n,m} D_- A_i^{(n)} [A_i^{(m)}, \bar{X}^{(n+m)}]
    \bigg) \ .
\end{align}
Here it is the non-relativistic kinetic term for the $A_i^{(n)}$ fields that diverges as we take $R_+\to 0$. However, unlike the divergence for $A_-^{(n)}$ this is not a squared quantity and there can be cancellations between divergent terms that render the final result finite. We will leave this term for the moment and come back to it momentarily. The finite part of the term becomes
\begin{align} \nonumber
    \cal{L}_2 = \tr \bigg(& - F_{-i} D_i X^{(0)} + \sum_{n\neq0} \Big( i F_{-i} [A_i^{(n)}, \bar{X}^{(n)}] - D_- A_i^{(n)} D_i \bar{X}^{(n)} \Big) \\
    &+ i \sum_{n,m\neq0} D_- A_i^{(n)} [A_i^{(m)}, \bar{X}^{(n+m)}] \bigg) \ .
\end{align}
Note that all field strengths and covariant derivatives on the right only include the zero-modes. The final term in the Yang-Mills action is
\begin{align} \nonumber
    \tr \int \frac{dx^+}{2\pi R_+} \, F_{ij} F_{ij} = \tr \bigg(&
    F_{ij} F_{ij} - 2i F_{ij} \sum_{n\neq0} [A_i^{(n)}, \bar{A}_j^{(n)}] + \sum_{n\neq 0} \abs{D_i A_j^{(n)} - D_j A_i^{(n)} }^2 \\ \nonumber
    &- 2i \sum_{\substack{n,m,p\neq 0 \\ n+m+p=0}} \brac{D_i A_j^{(n)} - D_j A_i^{(n)}} [A_i^{(m)}, A_j^{(p)}] \\
    &- \sum_{\substack{n,m,p,q\neq0 \\ n+m+p+q=0}} [A_i^{(n)}, A_j^{(m)}] [A_i^{(p)}, A_j^{(q)}]
    \bigg) \equiv \cal{L}_3 \ ,
\end{align}
which we see is finite as we take $R_+\to0$. 

We can do the same for the scalar fields. The kinetic term is
\begin{align}
    \tr \int \frac{dx^+}{2\pi R_+} \, D_+ Y^M D_- Y^M = \tr \bigg( 
    \sum_n \frac{i n}{R_+} \bar{Y}^M_{(n)} D_- Y^M_{(n)}  + i \sum_{n,m} D_- Y^M_{(n)} [Y^M_{(m)}, \bar{X}^{(n+m)}]
    \bigg) \ ,
\end{align}
so we see that we have an almost identical divergence to that for the $\{A_i^{(n)}\}$ fields arising from the kinetic term. Since these appear at the same order in $R_+^{-1}$, the most general constraint we can impose to render the theory finite is
\begin{align} \nonumber
    \cal{G}[\{A_i^{(n)}\},\{Y^M_{(n)}\}] \equiv \sum_{n=1}^{\infty} &\tr \int dx^- d^4x \bigg(D_- A_i^{(n)} \bar{A}_i^{(n)} - A_i^{(n)} D_- \bar{A}_i^{(n)} \\
    &+ D_- Y^M_{(n)} \bar{Y}^M_{(n)} -  Y^M_{(n)} D_- \bar{Y}^M_{(n)}\bigg) = 0 \ .
\end{align}
This allows for more general solutions than $A_i^{(n)} = Y^M_{(n)}= 0$, so the higher Fourier modes do not decouple. However, it is not clear whether there is a way to deal with such a complicated constraint. The finite parts of the term become
\begin{equation}
    \tr \int \frac{dx^+}{2\pi R_+} \, D_+ Y^M D_- Y^M \to \cal{L}_4 =  i\tr\sum_{n,m} D_- Y^M_{(n)} [Y^M_{(m)}, \bar{X}^{(n+m)}] \ .
\end{equation}

A quick calculation shows that there are no divergent terms from either the spatial gradient or interaction terms, with the Lagrangian
\begin{align}
    \cal{L}_5 = \tr \Bigg( & \nonumber
    - D_i Y^M_{(0)} D_i Y^M_{(0)} + \inv{2} [Y^M_{(0)}, Y^N_{(0)}] [Y^M_{(0)}, Y^N_{(0)}] + \sum_{n\neq0} \bigg( 2i D_i Y^M_{(0)} [A_i^{(n)}, \bar{Y}_{(n)}^M] \\ \nonumber
    &- \abs{D_i Y^M_{(n)} - i [A_i^{(n)}, Y^M_{(0)}]}^2  + 2 [Y^M_{(0)}, Y^N_{(n)} ] [Y^M_{(0)}, \bar{Y}^N_{(n)} ] - 2 [Y^M_{(0)}, Y^N_{(n)} ] [Y^N_{(0)}, \bar{Y}^M_{(n)} ] \\ \nonumber
    &+ 2 [Y^M_{(0)}, Y^N_{(0)}] [Y^M_{(n)}, \bar{Y}^N_{(n)}] \bigg) + \sum_{\substack{n,m,p\neq0 \\ n+m+p=0}} \bigg(
    2i \brac{D_i Y^M_{(n)} - i [A_i^{(n)}, Y_{(0)}^M]} [A_i^{(m)}, Y^M_{(p)}] \\ \nonumber
    &+ 2 [Y^M_{(0)}, Y^N_{(n)}] [Y^M_{(m)}, Y^N_{(p)}]
    \bigg) + \sum_{\substack{n,m,p,q \neq 0 \\ n+m+p+q=0}} \bigg(
    [A_i^{(n)}, Y^M_{(m)}] [A_i^{(p)}, Y^M_{(q)}] \\
    &+\inv{2} [Y^M_{(n)}, Y^N_{(m)}] [Y^M_{(p)}, Y^N_{(q)}]
    \bigg) \Bigg)
\end{align}
after integrating over $x^+$.

Putting this all together, the bosonic part of the path integral as we take $R$ and $R_+$ to zero with $k$ held fixed is (relabelling $x^-$ to $t$)
\begin{subequations}
\begin{align}
    Z &= \int DA_- \prod_n \brac{D A_i^{(n)} DX^{(n)} DY^M_{(n)}} \, \delta[\cal{G}] e^{i S_B} \ , \\
    S_B &= \frac{k}{4\pi} \tr \int dt d^3x \, \sum_{p=1}^5 \cal{L}_p \ .
\end{align}
\end{subequations}

\printbibliography

\end{document}